\documentclass[12pt]{article}
\usepackage{times} 
\usepackage{subfigure} 
\usepackage{amsmath} 
\usepackage{amsfonts} 
\usepackage{amsthm} 
\usepackage{graphicx}
\usepackage{latexsym}
\usepackage{fullpage}
\usepackage{delarray}
\usepackage{url}
\usepackage{verbatim}
\def\MyStretch{1.1}
\def\baselinestretch{\MyStretch}

\newcommand{\figdir}{./figs}

\newcommand{\beql}{\begin{equation}}
\newcommand{\eeql}{\end{equation}}
\newcommand{\beq}{\[}
\newcommand{\eeq}{\]}
\newcommand{\beqal}{\begin{eqnarray}}
\newcommand{\eeqal}{\end{eqnarray}}
\newcommand{\beqa}{\begin{eqnarray*}}
\newcommand{\eeqa}{\end{eqnarray*}}
\newcommand{\bfig}{\begin{figure}[htb!]\begin{center}}
\newcommand{\efig}{\end{center}\end{figure}}
\newcommand{\btab}{\begin{table}[htb!]\begin{center}}
\newcommand{\etab}{\end{center}\end{table}}
\newcommand{\btabs}{\begin{table*}[htb!]\begin{center}}
\newcommand{\etabs}{\end{center}\end{table*}}

\newtheorem{prop}{Proposition}

\newcommand{\urlBiBTeX}[1]{\url{#1}} 
\newcommand{\urlbibteX}[1]{\url{#1}} 

\title{\bf Providing Service Guarantees in High-Speed Switching Systems with Feedback Output Queuing\footnote{This paper is a 
revised and extended version of \cite{Firoiu02}.}}
\author{ 
\begin{tabular}{cccc} 
      Victor Firoiu
      & Xiaohui Zhang
& Emre G\a"{u}nd\a"{u}zhan & Nicolas Christin\footnote{Work done while visiting Nortel Networks.}\\
\multicolumn{3}{c}{\{vfiroiu,xiaohui,egunduzh\}@nortelnetworks.com} & christin@sims.berkeley.edu
\\
\multicolumn{3}{c}{Advanced Technology} & S.I.M.S.\\
\multicolumn{3}{c}{Nortel Networks} & UC Berkeley\\
\multicolumn{3}{c}{600 Technology Park} & 102 South Hall\\
\multicolumn{3}{c}{Billerica, MA 01821 USA} & Berkeley, CA 94720 USA
\end{tabular}
}

\begin{document}
\maketitle

\setcounter{footnote}{0}
\thispagestyle{empty}
\def\mynormality#1{\def\baselinestretch{#1}\small\normalsize}

\mynormality{1.0}

\vspace{0.5in}
\begin{abstract}
We consider the problem of providing service guarantees in a high-speed 
packet switch. As basic requirements, the switch should be scalable to  
high speeds per port, a large number of ports and a large number of     
traffic flows with independent guarantees. Existing scalable solutions  
are based on Virtual Output Queuing, which is computationally complex  
when required to provide service guarantees for a large number of       
flows.                                                                  

We present a novel architecture for packet switching that provides      
support for such service guarantees. A cost-effective fabric with       
small external speedup is combined with a feedback mechanism that       
enables the fabric to be virtually lossless, thus avoiding packet drops 
indiscriminate of flows. Through analysis and simulation, we show that  
this architecture provides accurate support for service guarantees, has 
low computational complexity and is scalable to very high port speeds.  
\end{abstract}
\noindent{\small \em
Keywords: Computer networks, Packet switching, Quality of service, Feedback
control, Congestion control.
}

\mynormality{1.5}

\newpage

\section{Introduction}

High speed communication between businesses has been a large share of
telecommunications market in recent years. This communication needs to
be of high quality, secure and reliable. Traditionally, these services
were provided using ATM and Frame Relay technologies, but at a premium
cost. Recent advances in traffic engineering and the advent of Voice
over IP technologies provide an opportunity to carry all enterprise
traffic (voice, streaming and non-real-time data) at a lower cost.
Virtual Private Networks (VPNs) \cite{Gleeson00-VPN-framework-IETF}
and Virtual Private LAN Services (VPLS) \cite{Waldemar-VPLS-req-IETF}
are two examples of such network services. A main requirement for such
services is to provide quality of service (QoS) guarantees. Interactive
media such as VoIP needs low delay and low loss, other traffic needs
minimum throughput guarantees.

In this paper we consider the problem of providing such guarantees in
a high-speed, cost-effective switch at the interface (edge) between
enterprise and service provider networks. At a minimum, the switch is
required to provide three types of service: Premium, Assured and Best
Effort \cite{Davie02-EFPHB-IETF},\cite{Heinanen-AFPHB-1999}. Premium
service provides low loss and small delay for a flow sending within
a pre-determined rate limit (anything above the limit is discarded).
Assured service guarantees delivery for traffic within a limit, but
allows and forwards extra traffic within a higher limit if transmit
opportunities are available.

A provider edge switch is required to differentiate between traffic
from different customers (here called flows) and provide separate
guarantees to each flow. A requirement is to support a large number
(in the order of hundreds or even thousands) of such flow guarantees
per port, where each port must support speeds in the order of several
Gbps. Traffic from one customer (flow) can enter through one or multiple
ingress ports and exit through one or multiple ports. On the other hand,
to come up with practical solutions, we assume that the
provided service guarantees only need to be enforced over timescales in
the order of a few milliseconds, which is enough for most applications,
thereby alleviating the traditional requirement that service guarantees
have to be enforced over timescales as small as a single packet
transmission time. We consider the problem of providing 1-to-1 and
N-to-1 services (or ``Pipe'' and ``Funnel scope'' as defined in
\cite{Goderis02-SLS-IETF}), as 1-to-N and N-to-N can be provided as
combinations of services of the first two kinds. In the case of Assured
N-to-1 service, it is also desirable to provide a fair distribution of
service among the N components of the flow.

Current state-of-the-art switch architectures are based on Virtual
Output Queuing (VOQ), which requires a fabric speedup $s\geq 2$ and a
matching algorithm to find which packets are sent into the fabric at
each fabric cycle. However, realizing a speed-up of $s \geq 2$ may be impractical at very high line speeds ($>10$~Gbps) given the limitations on memory access speeds. Furthermore, even though some of the VOQ architectures can support
service guarantees, a major problem is that the matching algorithms have
high complexity, are run at each fabric cycle, and all virtual output queues 
at all input
lines in the system need to participate in a centralized algorithm \cite{Nong-provision-commmag-2000}.

To provide a low-complexity switch architecture that fulfills the above
requirements, we observe that the main cause for high complexity in
current architecture resides in the necessity of addressing congestion at an
output line. Short term congestion can be absorbed by buffers, whereas
long term congestion results in packet loss. We also observe that
many measurement studies (for example \cite{Creary00-trends-CAIDA})
have shown that traffic in the Internet is dominated 
by the TCP protocol, which accounts for about 90\% of all traffic. 
A salient feature of TCP is that packet
transmission is controlled by a congestion avoidance algorithm
\cite{Jacobson:cong-avoid}, \cite{Stevens-rfc2001}. As an effect, the
average sending rate of a TCP flow is a decreasing function of drop
probability and of round trip time (see \cite{TCPchar-TON-00} for a
quantitative evaluation of this function). In practice, TCP flows have
a stable (long-term) operation at when the drop probability is between 0 and 0.1, corresponding to loss rates less than 10\%, and very
rarely operate above $0.2$ \cite{TCPchar-TON-00}. Heavy long-term
congestion that results in a drop probability above $0.2$ can be produced
by non-TCP (and more generally, non-congestion-controlled) traffic such as
multimedia traffic over UDP.

Our proposed architecture, named ``Feedback Output Queuing'' (FOQ),
exploits these observations by efficiently supporting fast fabrics with
relatively slow output memory interfaces and hence a small effective
speedup. For example, a speedup of $1.25$ at the fabric-to-line
interface is sufficient to maintain an output drop probability up to
$0.2$ for traffic flows  fully utilizing this interface. For higher levels of
long-term congestion (e.g., drop probability above $0.2$), the
FOQ architecture uses a feedback mechanism to reducing the traffic volume
before it enters the switch fabric. This FOQ mechanism provides support
for the Assured service, 1-to-1 and N-to-1 scope.

As far as Premium traffic is concerned, given that rate guarantees
are ensured to be within switch capacity by some admission control
procedure, policing Premium traffic at its guaranteed rate at the
ingress guarantees that Premium traffic cannot create congestion in the
absence of other types of traffic. Thus, Premium service can be provided
through a simple priority scheduling in OUT ports and fabric, bypassing
the FOQ mechanism.

In the following we show through analysis and simulation studies that
the proposed FOQ architecture can alleviate congestion at the output
lines of an output queued switch with slow output memory interface,
and can thus provide deterministic QoS guarantees. FOQ requires
only a modest speedup (e.g., 1.3) at the output interface of the
switch. The congestion control algorithm in the FOQ architecture is
fully parallelized at the input and output lines, requiring $O(1)$
complexity at each input and output line. This low complexity enables
implementation of the FOQ architecture at very high line rates ($>10$~Gbps).

The rest of the paper is organized as follows. In the next section we
discuss the related work in more details. Then, we give a detailed
description of the FOQ architecture in Section~\ref{sec-FOQ-arch}
In Section~\ref{sec-PI-GB-model} we develop an analytical
model for FOQ, based on a PI controller, and analyze its
performance under step-shaped traffic bursts, before introducing
a quantized version of a PI controller. We present our simulation
results in Section~\ref{sec-simul-burst}, and
conclude the paper with a comparison between FOQ and VOQ in
Section~\ref{sec-conclusion}.
\section{Related Work}
Several switch
architectures with QoS capabilities have been proposed in the
literature, with particular advantages and shortcomings.

An early architecture is Output Queuing (OQ). An OQ switch
having $N$ inputs and $N$ outputs with each line of speed $c$
bits/second requires a switching fabric of speed $Nc$, i.e., a
speedup $s=N$.  In this case, no congestion occurs at the inputs or at
the fabric, only at the output lines.  To manage congestion and
provide QoS support, a set of queues and a scheduling mechanism is
implemented at each output.  The main advantage of this architecture is that
it can provide QoS support with simple mechanisms of queuing
and scheduling, but the main problem is that the fabric speedup of $N$
can be impractical. In fact current technology enables fast interconnection
networks operating at current high speed line rates and with typical number of lines
(for example $c=10$~Gbps and $N=16$), but writing the packets coming
out of the interconnection network into output buffers at high speeds remains
a problem. In other words, although the fabric may have an internal
speedup of $N$, the effective speedup seen at an output buffer is limited by
the memory write speed which is usually much less.

An alternative to OQ is Virtual Output Queuing
(VOQ) \cite{Anderson-hispeedswitch-tocs-1993},
\cite{McKeown-comparison-CNIS-1998}, which requires a smaller fabric
speedup, such as $s$ in the range 
between 2 and 4. Unlike OQ, VOQ requires a matching
algorithm to find which packets will be sent into the fabric at
each fabric cycle. There are quite a few such algorithms proposed
in the literature, which are based on Parallel Iterative Matching,
Time Slot Assignement, Maximal Matching, or Stable Matching (see
\cite{Nong-provision-commmag-2000} and references therein). Some of
these algorithms can also support service guarantees. The advantage of VOQ
is its ability to switch high speed lines with low fabric speedup.
However its main problem is that the matching algorithms are complex
($O(M^2N^2)$ where $M$ is the number of independent service guarantees per
port, $N$ is the number of ports), have to be run at each fabric cycle,
and all VOQs at all input lines in the system need to participate in a
centralized algorithm. We note that Output Queued switches can also be
perfectly emulated by Combined Input-Output Queued (CIOQ) switches with
a speed-up $s\geq 2$ \cite{McKeown99a}. Unfortunately, the arbitration
algorithm has a computational complexity of $O(N^2)$, which can be
reduced to $O(N)$, but in that case, the space complexity becomes linear
in the number of cells in the switch. Therefore, emulating an OQ switch
by a CIOQ switch or a VOQ switch appears to have limited scalability.

In recent years, these potential scalability concerns have been 
addressed by
implementing a very small number of independent service guarantees.
Under the Differentiated Services framework \cite{dsarch-rfc},
flows are aggregated in $M=6$ classes, and service guarantees are offered 
for classes. The downside is that the realized QoS per flow has a
lower level of assurance (higher probability of violating the desired service 
level) than the QoS per aggregate \cite{GuerinPla01},
\cite{Xu-individual-2002}. Moreover, recently proposed VPN and VLAN
services \cite{Rosen-L2VPN-2001}, \cite{Carugi-PPVPNreq-2002}  require
per-VPN or VLAN QoS guarantees. All the above are arguments in favor of
implemeting a number of independent service guarantees per port much larger
than six.

More recent proposals \cite{Karr00-reduced-HotI} decrease the time
interval between two runs of the matching algorithm, but with a
tradeoff in increased burstiness and additional scheduling algorithms
for mitigating unbounded delays. Moreover, the service presented in
\cite{Karr00-reduced-HotI} is of type Premium 1-to-1, but cannot provide
Assured N-to-1 service.

Last, similar to the FOQ architecture proposed in this paper, the IBM
Prizma switch architecture \cite{Prizma} uses a shared memory, and no
centralized arbitration algorithm. However, Prizma relies on on-off
flow control while the feedback scheme proposed in the present paper
dynamically controls the amount of traffic admitted into the fabric, and
FOQ feedback is based on the state of the output queues, while Prizma
relies on the state of internal switch queues. Both the origin of the
information and the dynamic control of the drop level lead us to believe
that FOQ can use the capacity available in the switch more efficiently.
\section{Feedback Output Queuing Architecture} 
\label{sec-FOQ-arch}

\begin{figure}
\begin{center}
\includegraphics[width=5in]{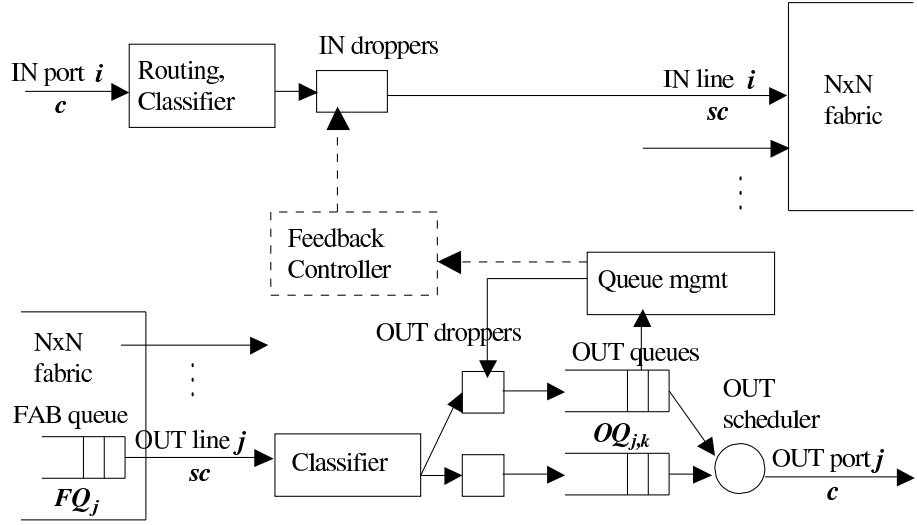}
\end{center}
\caption{Detailed FOQ switch architecture}
\label{fig-foq-arch}
\end{figure}

We consider a switch as in Figure~\ref{fig-foq-arch} with a             
fabric having internal speedup of $N$ and an internal buffer            
capability.\footnote{This fabric has a cost-effective implementation    
using shared memory technology. The case of zero/small memory fabric    
with no/small internal speedup is a separate problem, and we report     
our study elsewhere.} We also assume that the fabric has one or a       
very small number of queues per port. In the following we present an    
architecture for providing per-flow service guarantees where the number of  
flows per port $M$ is large, that is, $M\gg 1$.

Packets enter through a set of $N$~input ports of speed $c$. As a       
packet is received at port~$i$, a destination port $j$ is determined by 
a routing module, its QoS flow~$k$ is determined by a classifier and   
an IN dropper determines if the packet is discarded. If not discarded,  
the packet is transmitted to the fabric through a line of speed $sc$.   
We assume a fabric with internal speed of $Nsc$, i.e., at each fabric   
cycle one packet from each IN line can be moved to an OUT line while    
sustaining speeds of $sc$ from all IN lines. Multiple (up to $N$)       
packets can be received at an OUT line in one cycle, and in that case   
the packets are placed in a fabric queue $FQ_j$ corresponding to the    
destination line $j$.

Packets are forwarded by the OUT line $j$ at speed $sc$, separated
into OUT queues $\{OQ_{j,k}\}_{k}$ based on their QoS flow, and
scheduled for transmission to OUT port $j$ of speed $c$.  The OUT
scheduling implements various service guarantees such as priority, minimum
rate guarantee, maximum rate limit, maximum delay guarantee.  This OUT
scheduling results in a certain service rate (in general variable in
time) for each OUT queue.

If traffic to $OQ_{j,k}$ has a rate higher than the current service
rate of flow~$k$, packets accumulate in this queue and some of them
may be dropped by a queue management mechanism such as drop-tail or
RED (see \cite{Floyd-red} for details).  If the traffic to all queues at
OUT line $j$ amounts to an aggregate rate above $sc$, then packets
accummulate at the fabric queue $FQ_j$.  If this situation persists,
$FQ_j$ fills and packets get dropped in the fabric.  In this case, QoS
guarantees for some flow~$k$ may be violated since fabric drops 
do not discriminate between different flows.

We define the {\em relative congestion} at a queue
\beql
C=1-\frac{r_O}{r_I}
\label{eq-C}
\eeql
where $r_I$ and $r_O$ are traffic rates input to and output from the
queue respectively.  It is easy to see that, as long as the traffic
coming out of OUT line $j$ is such that the relative congestion
$C_{j,k}$ at each queue $\{OQ_{j,k}\}_{k}$ is below a threshold
$d_{max}<1-1/s$, and the OUT port $j$ is utilized at its full capacity
$c$, then the traffic throughput at the interface of fabric to OUT
line $j$ is below $sc$, and thus there is no congestion at that
interface and no fabric drop.

In the FOQ architecture, a feedback mechanism is introduced to control  
the relative congestion at each OUT queue below a threshold. When the   
relative congestion at an OUT queue increases, the feedback mechanism   
instructs the input modules to drop a part of the traffic destined to   
this queue. By keeping the traffic below a congestion threshold, the    
fabric drop is avoided. Thus, packet are dropped only from those flows  
that create congestion, and the QoS guarantees are provided to all      
flows as configured.                                                    

It is worth noting that the flows having packets dropped at ingress by
FOQ would have packets dropped in the same amount at egress in the
case of an ideal Output Queuing with speedup of $N$.  Thus, FOQ
reduces the demand of fabric throughput by eliminating the need for
forwarding packets that are later discarded.

\paragraph{Realizations of FOQ} 
We next consider options for a practical realization of the FOQ         
architecture. More precisely, we consider implementations of FOQ as     
a discrete feedback control system. A certain measure of congestion     
is sampled at intervals of duration $T$ at each OUT queue. A control    
algorithm computes a drop indication based on the last sample and an    
internal state, and transmits it to all IN modules. There, packets of   
the indicated class are randomly dropped with a probability that is a   
function of the drop indication.

We have several ways to measure the congestion at a queue.  A simple
method is to compute the average drop probability at the queue during
the sampling interval:
\beq
DropProb(T)=DroppedPkts(T)/InPkts(T) \ .
\eeq
Another measure is the relative congestion
during the interval $T$, similar to (\ref{eq-C}):
\beq
RelCong(T)=1-OutPkts(T)/InPkts(T) \ .
\eeq
Observe that, unlike the drop probability, the relative congestion
takes into account the variation of the queue size during $T$.  Since
the FOQ objective is to keep the traffic rate at the fabric interface
below a critical level, it is apparent that the relative congestion is
more effective in controlling that traffic rate.  This is confirmed by
the model in Section~\ref{sec-PI-GB-model} and the simulation in
Section~\ref{sec-simul-burst}. 

We consider a discrete Proportional-Integrator (PI)
\cite{Franklin-digitalcontrol-1998} for the feedback control
algorithm.  In Section~\ref{sec-PI-GB-model} we derive its
configuration from stability conditons.  The PI algorithm outputs a
value of drop probability between $0$ and $1$ transmitted to the IN
droppers every interval.

An implementation issue is the data rate of feedback transmission.      
Considering $K$ classes at each of the $N$ OUT ports and that the       
drop information is coded in $F$ bits, the total feedback data rate     
is $KNF/T$. For example, for $K=1000$, $N=32$, $F=8$, $T=1$~ms, the     
feedback data rate is $256$~Mb/s. It is possible to reduce this rate by 
reducing the precision of the feedback data, and thus its encoding. In  
an extreme case, the feedback has three values: increase, decrease or   
keep same drop level. All IN modules use this indication in conjunction 
with a pre-defined table of drop levels. We call this the ``Gear-Box    
algorithm'' (GB), model it in Section~\ref{sec-PI-GB-model} and show    
its performance in Section~\ref{sec-simul-burst}.                       

\section{A Control Theoretical Model for the GB Algorithm} 
\label{sec-PI-GB-model}

In this section we develop an analytical model for the FOQ architecture
by a control theoretical approach. In our analysis, we use a classical
discrete PI controller to adjust the drop rate of each flow. We simplify
our analysis by assuming only a single flow at first, and later discuss
how and under what conditions our results may apply to the general
multi-flow case. We also assume in our analysis that there is no
limitation to the capacity of the feedback channel in the system. We
then show that an efficient algorithm for limited-capacity feedback
channels can be obtained by quantizing the control decisions of the PI
controller, which we call the Gear Box algorithm. 

\begin{figure}
\begin{center}
\includegraphics[width=0.6\textwidth]{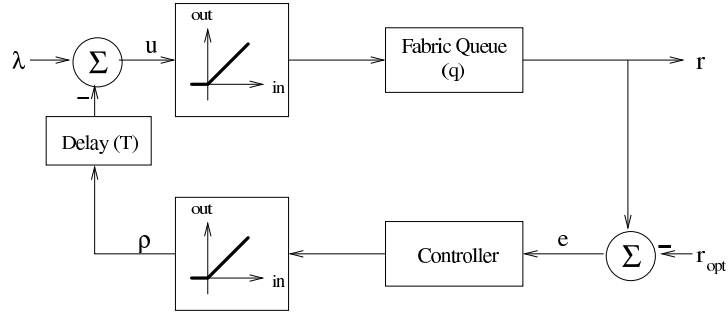}
\caption{\label{fig-controller} FOQ architecture.}
\end{center}
\end{figure}

The basic control structure at a particular OUT port $j$ and for a
particular flow $k$ is shown in Figure~\ref{fig-controller}. If there
are a total of $K$ flows in each OUT port, then each OUT port has $K$ such
controllers. All variables we use in this section are for the aggregate
traffic in flow $k$ originating from all IN ports and destined to OUT
port $j$, unless we note otherwise (i.e., we don't use the subscript
$(j,k)$ for notational convenience).
$\lambda$ is the total arrival rate for traffic destined for the OUT queue
$OQ_{j,k}$. A total portion, $\rho$, of the arriving traffic is
dropped at the IN droppers, and the surviving portion goes into the fabric
queue $FQ_j$ at a rate $u=\lambda-\rho$. This traffic shares the fabric queue
with other traffic destined to OUT line $j$, and then it is delivered to OUT
dropper $(j,k)$ at a rate $r$. In the analysis we assume the fabric queue is
sufficiently large, so that there are no drops due to queue overflow.

The total drop rate, $\rho$, is adjusted by a controller (how $\rho$
is distributed among the $N$ IN droppers is not relevant for this analysis; we
explain how we implement the actual drop mechanism in the next section). The
purpose of the controller is to keep the fabric output rate for packets
destined to $OQ_{j,k}$ at a desired level, $r_{opt}$. The desired rate can be
chosen according to the current rate out of $OQ_{j,k}$
$$
r_{opt}=\alpha sr_{O(j,k)},
$$
where $\alpha$ is a constant smaller than but close to 1. In this way the
desired rate will be close to the capacity, $sc$, of fabric output line when
the OUT queue $OQ_{j,k}$ is the only busy queue and utilizing the entire speed
of port $j$. Furthermore it will be reduced in proportion to the service rate
of $OQ_{j,k}$ when multiple OUT queues are contending for the OUT port. The
two nonlinearities in the figure simply state that the drop rate can not be
negative or greater than the arrival rate $\lambda$. In our analysis we assume
that the controller is operating in the linear region, and ignore the
nonlinearities.

The delay $T$ between the output of the controller and the arrival rate
models a zero-order hold at the controller output. The controller operates on
time-average of the error signal taken over an interval $T$, rather than the
signal itself, and modifies its output only at intervals of $T$. In the rest
of this section we denote the time-average of a signal $x(t)$ over the period
$T$ by the discrete notation $x[n]$. For example the time-average of the
fabric output rate is given by
$$
r[n]=\frac{1}{T}\int_{nT}^{(n+1)T}r(t)dt.
$$
When the system is in steady state, the amount of traffic, $q$, in the fabric
queue destined to $OQ_{j,k}$ does not change significantly during the interval
$T$. Therefore, we can approximate the average fabric output rate by
\begin{eqnarray}
r[n] & \approx & \frac{1}{T}\int_{nT}^{(n+1)T}u(t)dt\nonumber \\
\label{out_rate}
& = & \lambda[n]-\rho[n-1].
\end{eqnarray}
For a discrete PI controller the drop rate for the next interval is
calculated using the error between the average fabric output rate, $r[n]$,
and the desired fabric output rate, $r_{opt}[n]$, 
\begin{eqnarray}
\rho[n] & = & Ke[n]+K_I\sum_{m=0}^{n}e[m]\nonumber\\
\nonumber
& = & K(r[n]-r_{opt}[n]) \\
& & +K_I\left(\sum_{m=0}^{n}r[m]-\sum_{m=0}^{n}r_{opt}[m]\right)\nonumber.
\end{eqnarray}

We can now investigate the step response of the system, setting
$\lambda[n]=\lambda_0$ and $r_{opt}[n]=r_{opt}$ for $n\geq 0$, for the case of a single flow. The magnitude of the
arrival rate can in general be larger than the maximum fabric output rate,
i.e., $\lambda_0>sc$. In this case the fabric output will be constant at
$r[n]=sc$ for an initial period $0\leq n<N_0$. During this period the fabric queue will always be
non-empty and the controller can not sense the actual magnitude of the
arrival rate. Therefore the controller output will increase linearly,
$$
\rho[n] = K(sc-r_{opt})+(n+1)K_I(sc-r_{opt}).
$$
The fabric queue size, measured at the end of each period, will increase until the
drop rate reaches $\lambda_0-sc$ and then decrease back to zero
\begin{eqnarray}
q_n & = & T\sum_{m=0}^{n}(\lambda_0-sc-\rho[m-1]) \nonumber \\
 & = & T[(n+1)(\lambda_0-sc)-nK(sc-r_{opt})\nonumber \\
\label{queuesize}
 & & -\frac{n(n+1)}{2}K_I(sc-r_{opt})].
\end{eqnarray}
The duration of this initial period, $N_0$, and the maximum queue size can
easily be calculated from this quadratic equation setting $q_{N_0-1}=0$. To find
the behavior of the system for $n\geq N_0$ we use a new time axis,
$n^{\prime}=n-N_0$, with an initial condition for the accumulator memory
\begin{eqnarray}
\nonumber
\rho[n^\prime] & = & K(r[n^\prime]-r_{opt}[n^\prime]) \\
 & & + K_I\left(\sum_{m=0}^{n^\prime}r[m]-\sum_{m=0}^{n^\prime}r_{opt}[m]\right)+S_{N_0}\nonumber\\
\label{rhonew}
\end{eqnarray}
where
$$
S_{N_0}=K_I N_0(sc-r_{opt}).
$$

Equations~(\ref{out_rate}) and (\ref{rhonew}) describe a closed-loop
control system. We show in the appendix that the two poles of this
system are at
$$
z_1=-\frac{K+K_I-1}{2}+\frac{1}{2}\sqrt{(K+K_I-1)^2+4K}
$$
$$
z_2=-\frac{K+K_I-1}{2}-\frac{1}{2}\sqrt{(K+K_I-1)^2+4K}.
$$
It follows that we have the stability condition given by the proposition below.
\begin{prop}
The closed-loop system described by (\ref{out_rate}) and (\ref{rhonew})  is
stable iff
\begin{equation}
0<K_I<2(1-K).
\label{eq:prop1}
\end{equation}
\end{prop}
\begin{proof} If $K+K_I>1$ then $|z_2|>|z_1|$, and both poles are inside the
unit circle iff
$$
K+K_I-1+\sqrt{(K+K_I-1)^2+4K}<2,
$$
which yields
$$
K+\frac{K_I}{2}<1.
$$
On the other hand if $K+K_I<1$ then $|z_2|<|z_1|$, and both poles are
inside the unit circle iff
$$
-(K+K_I-1)+\sqrt{(K+K_I-1)^2+4K}<2,
$$
which yields 
$$
K_I>0.
$$
Combining the two cases gives the condition for stability.
\end{proof}

In the appendix we solve the system with the stability
condition~(\ref{eq:prop1}, and show that the controller output is given
by
\begin{equation}
\label{rho_result}
\rho[n]=\left\{\begin{array}{ll}
[K+(n+1)K_I](sc-r_{opt}),
& n< N_0 \\
D(1-A_1z_1^{n-N_0}+A_2z_2^{n-N_0}),
& n\geq N_0 \end{array}\right.
\end{equation}
where
$$
A_1=\frac{z_1^2-\frac{S_{N_0}}{D}z_1}{z_1-z_2},
$$
$$
A_2=\frac{z_2^2-\frac{S_{N_0}}{D}z_2}{z_1-z_2},
$$
and
$$
D=\lambda_0-r_{opt}
$$
is the difference between the arrival and the desired rates. We observe that
after the initial linear increase, the drop rate approaches exponentially
to the difference between the arrival and the desired rates. Furthermore,
since the absolute value of the negative pole is relatively larger for
$K_I>1-K$, the system
 will show more oscillatory behavior in this case
compared to the $K_I<1-K$ case.

\paragraph{Multiple flows}
When there are multiple flows, the analysis for the initial period $(n<N_0)$
needs 
to be updated. Let $v$ be the total rate of the traffic
that does not 
belong to flow $k$ but destined to port $j$. If the step size
 for flow $k$
is such that $\lambda +v>sc$ then for an initial period the average 
fabric
output rate for flow $k$ is approximately
$$
r[n]=sc\frac{u[n]}{v[n]+u[n]}.
$$
Since $r$ is not constant anymore, the previous results for the initial
period
 do not apply in general. However, once the transient is over and $u$
and $v$ are 
adjusted so that $u[n]+v[n]\leq sc$, the approximation
(\ref{out_rate}) holds, and the 
results for the single-flow case can be
used replacing $S_{N_0}$ by a new initial 
condition. We defer a detailed
analysis of the initial transient period for the 
multi-flow case to a
future study. However, in two cases, when $u$ or $v$ is negligible 
compared
to the other, the results for the single-flow case can be used with
 some
changes. If $u \gg v$, then $r[n]\approx sc$ and we can approximate
 the
multiple-flow case by the single-flow case. On the other hand, if $u \ll v$
then we can assume that $v$ is constant since the effect of the new traffic,
$u$ will be
 negligible. Therefore
$$
r[n]\approx sc\frac{u[n]}{v}=\sigma u[n]
$$
with $\sigma=sc/v$ during the initial period $n<N_0$. In this case $N_0$ is
defined by
$$
\lambda_0-\rho[N_0-1]+v=sc.
$$
For $n<N_0$ the drop rate can be calculated by replacing (\ref{out_rate}) with
$$
r[n]\approx\sigma (\lambda[n]-\rho[n-1]).
$$
The response for $n\geq N_0$ is still
 given by (\ref{rho_result}) but with a new
initial condition replacing $S_{N_0}$.

\paragraph{Quantized PI - the Gear Box algorithm}
A practical implementation of the discrete-time PI control described
above requires a few modifications to the control loop. The first
modification is related to how the bytes will actually be dropped at the
desired drop rate calculated by the controller. The drop rate has to be
divided fairly among the $N$ IN droppers. Furthermore it is well-known
that dropping consecutive packets may result in poor performance in the
affected flows. Therefore it is desirable to spread the drop rate to an
interval and to introduce some randomness into the drop process. For
these reasons we introduce a packet drop probability, $p[n]$, which is
updated at intervals of $T$ according to the desired drop rate and the
estimated average arrival rate,
\begin{equation}
\label{defp}
p[n]=\frac{\rho[n]}{\hat{\lambda}[n+1]}=\frac{(1-p[n-1])\rho[n]}{r[n]}.
\end{equation}

Note that here we used the fabric output rate divided by the admit
probability (i.e., $1-p[n-1]$) as an estimate of the next average
arrival rate. This is justified for the cases where the average arrival
rate is a slowly varying function relative to interval $T$ and the delay

The second modification to the feedback structure is related to the
constraint on the size of the feedback channel, which becomes a
limiting factor on the precision of the feedback signal at high speeds.
Our goal is to use only a finite number of drop probability
values, and to derive 
a controller that will have a similar performance with
the PI controller. For this
 purpose we expand (\ref{defp}) as
\begin{eqnarray}
p[n] &=& \frac{1}{\hat{\lambda}[n+1]}\left(Ke[n]+K_I\sum_{m=1}^{n}e[m]\right)\nonumber\\
&=& \frac{1}{\hat{\lambda}[n+1]}(Ke[n-1]+K_I\sum_{m=1}^{n-1}e[m]
+Ke[n]+K_Ie[n]-Ke[n-1])\nonumber \ .
\end{eqnarray}
Using again the assumption $\hat{\lambda}[n+1]\approx\hat{\lambda}[n]$, we can rewrite the above equation as
\begin{eqnarray}
p[n]&\approx& p[n-1]+\frac{1}{\hat{\lambda}[n+1]}(Ke[n]
+K_Ie[n]-Ke[n-1])\nonumber\\
&=& p[n-1]+\frac{(1-p[n-1])}{r[n]}(Ke[n]
+K_Ie[n]-Ke[n-1])\nonumber\\
&=& \left(1-\frac{(K+K_I)e[n]-Ke[n-1]}{r[n]}\right)p[n-1]
+\frac{(K+K_I)e[n]-Ke[n-1]}{r[n]} \ . \nonumber
\end{eqnarray}
Now, if 
we define
$$
\delta[n]=\frac{(K+K_I)e[n]-Ke[n-1]}{r[n]}
$$
then the update for the drop probability simply becomes
$$
p[n]=(1-\delta[n])p[n-1]+\delta[n].
$$
In order to use finite values of $p[n]$
 we quantize $\delta[n]$ to three
levels
\begin{equation}
\delta_q[n]=\left\{\begin{array}{ll}\beta & \delta[n] >\Delta_{\max}\\
0 & -\Delta_{\min}\leq\delta[n]\leq\Delta_{\max}\\
\frac{\beta}{\beta-1} & \delta[n]<-\Delta_{\min}
\end{array}\right.
\label{deltaq}
\end{equation}
Then the update for discrete probability values becomes
$$
p_q[n]=(1-\delta_q[n])p_q[n-1]+\delta_q[n],
$$
which can also be written as an update of admit probabilities as
$$
1 - p_q[n]=(1-\delta_q[n])(1-p_q[n-1]).
$$
If we set $K=0$, then (\ref{deltaq}) can also be expressed in terms of the
relative 
congestion $C[n]=1-r_O[n]/r[n]$ as
$$
\delta_q[n]=\left\{\begin{array}{ll}\beta & 
C[n] > d_{\max} \\
\frac{\beta}{\beta-1} & 
C[n] < d_{\min} \\
0 & \mbox{otherwise}
\end{array}\right. ,
$$
where
$$
d_{\max} = 1-\frac{1}{\alpha s}+\frac{\Delta_{max}}{\alpha sK_I},
$$
and
$$
d_{\min} = 1-\frac{1}{\alpha s}-\frac{\Delta_{min}}{\alpha sK_I}.
$$

We call the quantized mechanism with $K=0$ the {\em Gear Box (GB)} controller, since
there are only 
three possible actions: increase the drop probability,
decrease 
the drop probability, and no change. With the GB controller it is
sufficient to have a
 2-bit feedback signal every $T$ seconds. Furthermore the
different levels of the admit 
probabilities are the different powers of
$(1-\beta)$. Therefore the calculation at 
the IN droppers can be implemented
by storing
$$
P_k=1-(1-\beta)^k
$$
as a table in the memory and just updating a pointer to this table based on
the feedback 
signal.



To
increase the stability of the control loop, in our implementation 
of the GB algorithm, 
we
choose the value for $\beta$ such that the relative congestion after a
step increase or decrease in IN drop probability be equal. To find the 
value for $\beta$ that has this property, when note that  
when the relative congestion $C$ reaches $d_{\max}$, the drop step is increased, and the relative congestion immediately changes to a different value $C_{new,1}$. More precisely, if we have:
$$
C = 1-\frac{r_O}{r_I} = d_{\max} \ ,
$$
then $r_I$ changes to $r_{I, new} = r_I(1-\beta)$, so 
$$
C_{new,1} = 1-\frac{r_O}{r_I(1-\beta)} \ ,
$$
which can be rewritten as
$$
C_{new,1} = 1-\frac{1-d_{\max}}{1-\beta} \ .
$$
Likewise, when $C$ reaches $d_{\min}$, the drop step is decreased and the relative congestion immediately changes to a different value $C_{new,2}$. That is, 
$$
C = 1-\frac{r_O}{r_I} = d_{\min} \ ,
$$
has the effect of changing $r_I$ to $r_{I, new} = \frac{r_I}{(1-\beta)}$, yielding$$
C_{new,2} = 1-\frac{r_O(1-\beta)}{r_I} \ ,
$$
that is
$$
C_{new,2} = 1-(1-d_{\min})(1-\beta) \ ,
$$
and we want to have $C_{new,1} = C_{new,2}$. Hence,
$$
1-\frac{1-d_{\max}}{1-\beta} = 1-(1-d_{\min})(1-\beta) \ ,
$$
which reduces to
$$
\frac{1-d_{\max}}{1-d_{\min}} = (1-\beta)^2 \ ,
$$
giving finally
\begin{equation}
\beta=1-\sqrt{\frac{1-d_{max}}{1-d_{min}}}
\label{eq:hysteresis}
\end{equation}
as the value for $\beta$ such that the relative congestion after a
step increase or decrease in IN drop probability be equal.

\begin{figure}[t]
\begin{center}
\includegraphics[width=0.5\textwidth]{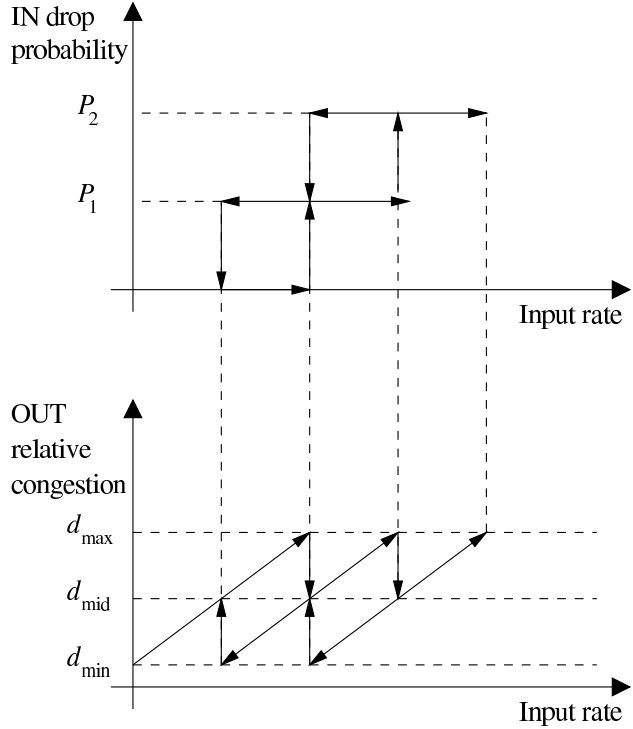}
\caption{\label{fig-GB-dynamics} FOQ dynamics and stability}
\end{center}
\end{figure}
We illustrate the behavior of the system when subject to the            
configuration of (\ref{eq:hysteresis}) in Figure~\ref{fig-GB-dynamics},  
where $d_{\mbox{\scriptsize
mid}} = 1 - \sqrt{(1-d_{\min})(1-d_{\max})}$. 
When the input rate increases such that the output relative congestion  
goes from $d_{\min}$ to $d_{\max}$, the input drop probability          
remains at the same level, and jumps to $P_1$ when the output           
relative congestion reaches $d_{\max}$. This jump in the input drop     
probability has the immediate effect of causing the output relative     
congestion to decrease to a value $d_{\mbox{\scriptsize 
mid}}$. Then, if the        
output relative congestion increases again to $d_{\max}$, the input     
drop probability remains at $P_1$ before jumping to $P_2$ when the      
output relative congestion reaches $d_{\max}$. Now, if the input drop   
probability is at $P_2$, and the relative congestion decreases from     
$d_{\mbox{\scriptsize 
mid}}$ to $d_{\min}$, the input drop probability remains      
at $P_2$, and jumps down to $P_1$ as soon as the relative congestion 
reaches $d_{\min}$. The decrease in the input drop probability from     
$P_2$ to $P_1$ immediately increases the output relative congestion to  
$d_{\mbox{\scriptsize 
mid}}$. 

As shown in      
Figure~\ref{fig-GB-dynamics}, this configuration has the key advantage  
of providing hysteresis to the GB control, by always trying to have the relative congestion come back to $d_{\mbox{\scriptsize mid}}$, 
thereby providing stability     
against small perturbations. We will use this configuration in our      
simulations presented in the following.                                 

\section{Simulation Experiments}
\label{sec-simul-burst}
The objective of this section is to present a set of experimental
results that illustrate the salient properties of FOQ. First, we describe
a relatively simple experiment with three classes of traffic and
constant-bit-rate (CBR) traffic, before presenting experimental results
gathered for a more realistic situation where traffic consists of a
large number of non-synchronized TCP sources.

\subsection{FOQ and Service Guarantees}
\begin{figure*}
\begin{center}
\shortstack{
        \includegraphics[width=0.45\textwidth]{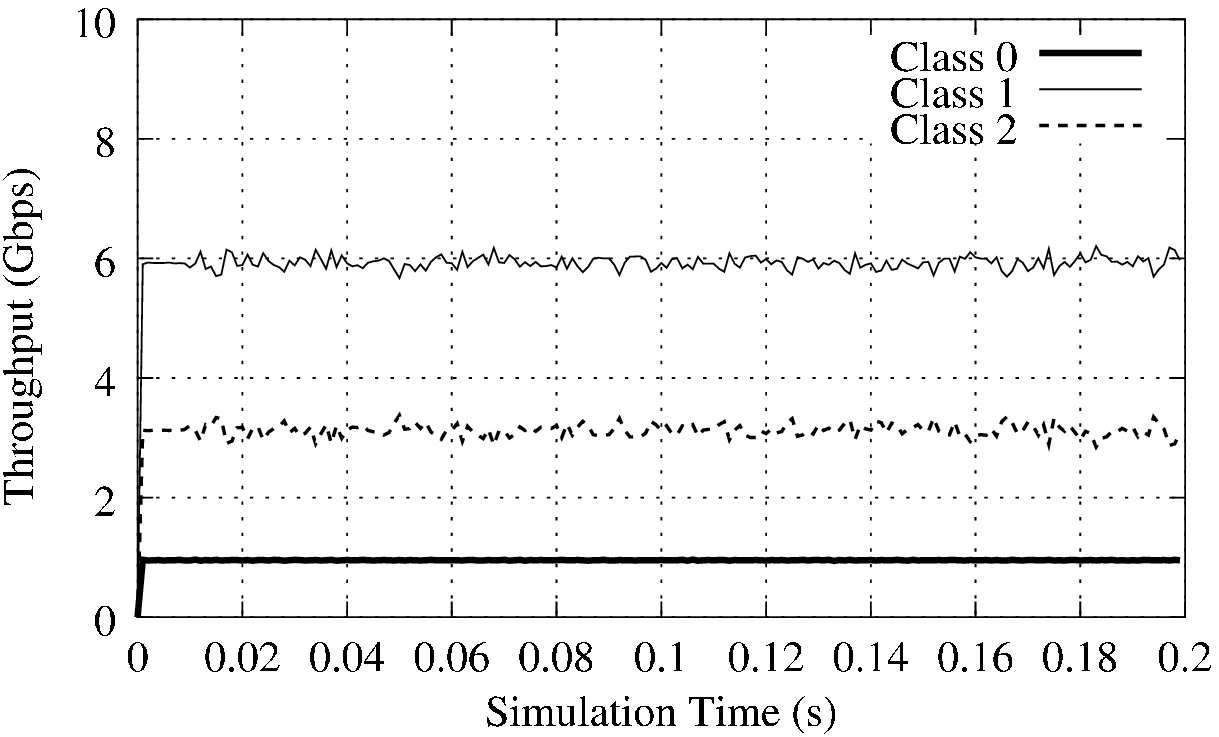}\\
        {\small (a) without FOQ}
}
\shortstack{
        \includegraphics[width=0.45\textwidth]{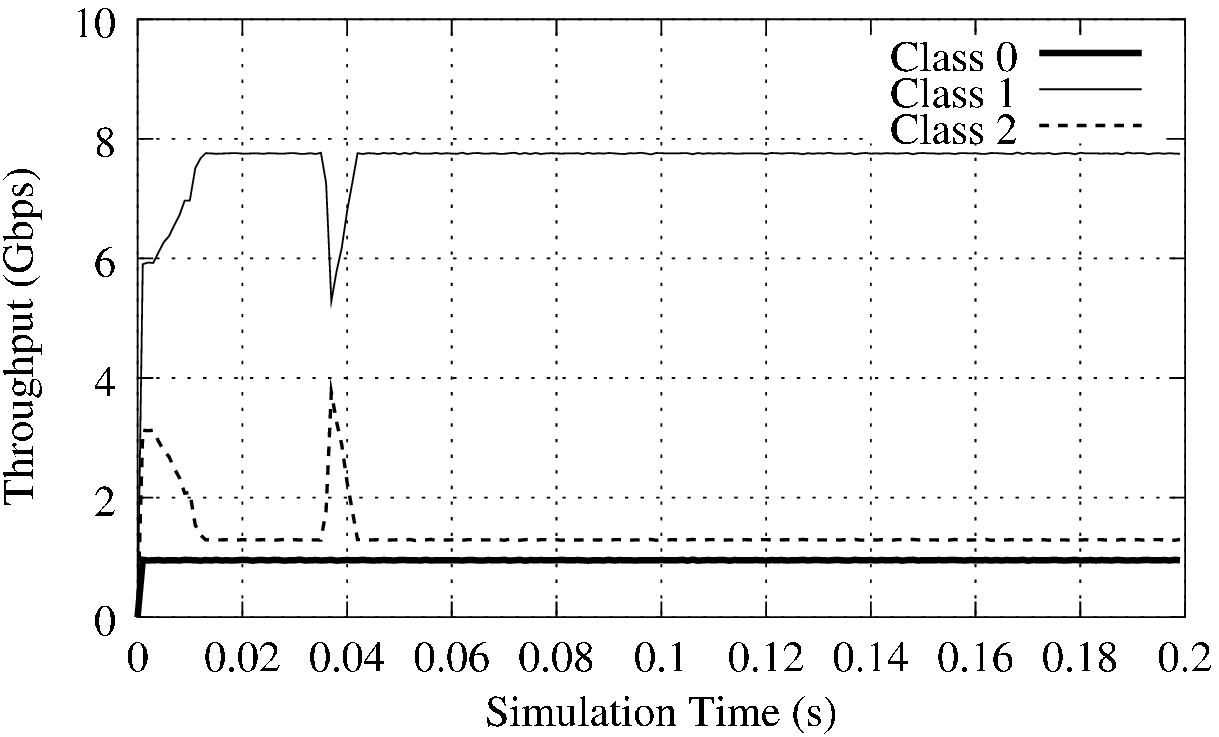}\\
        {\small (b) with FOQ}
}
\caption{Throughput plots}
\label{fig-exp1-thru}
\end{center}
\end{figure*}

\begin{figure*}
\begin{center}
\shortstack{
        \includegraphics[width=0.45\textwidth]{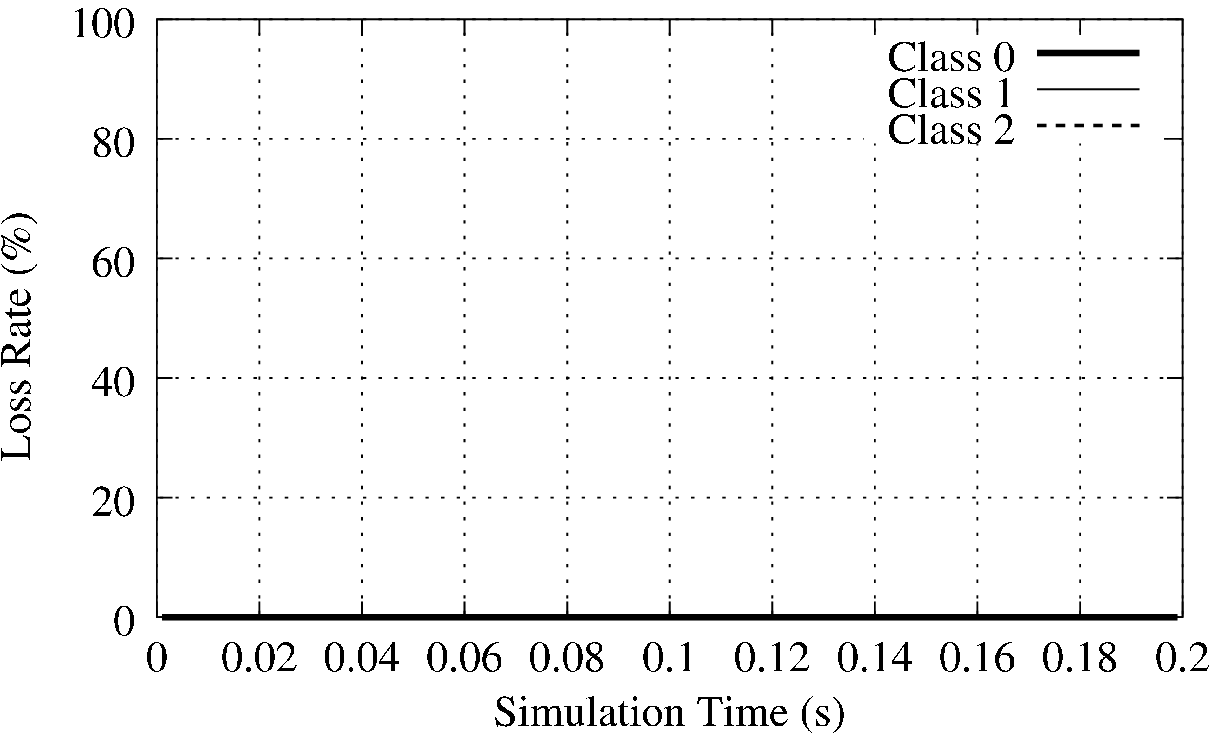}\\
        {\small (a) Input drop rate without FOQ}
}
\shortstack{
        \includegraphics[width=0.45\textwidth]{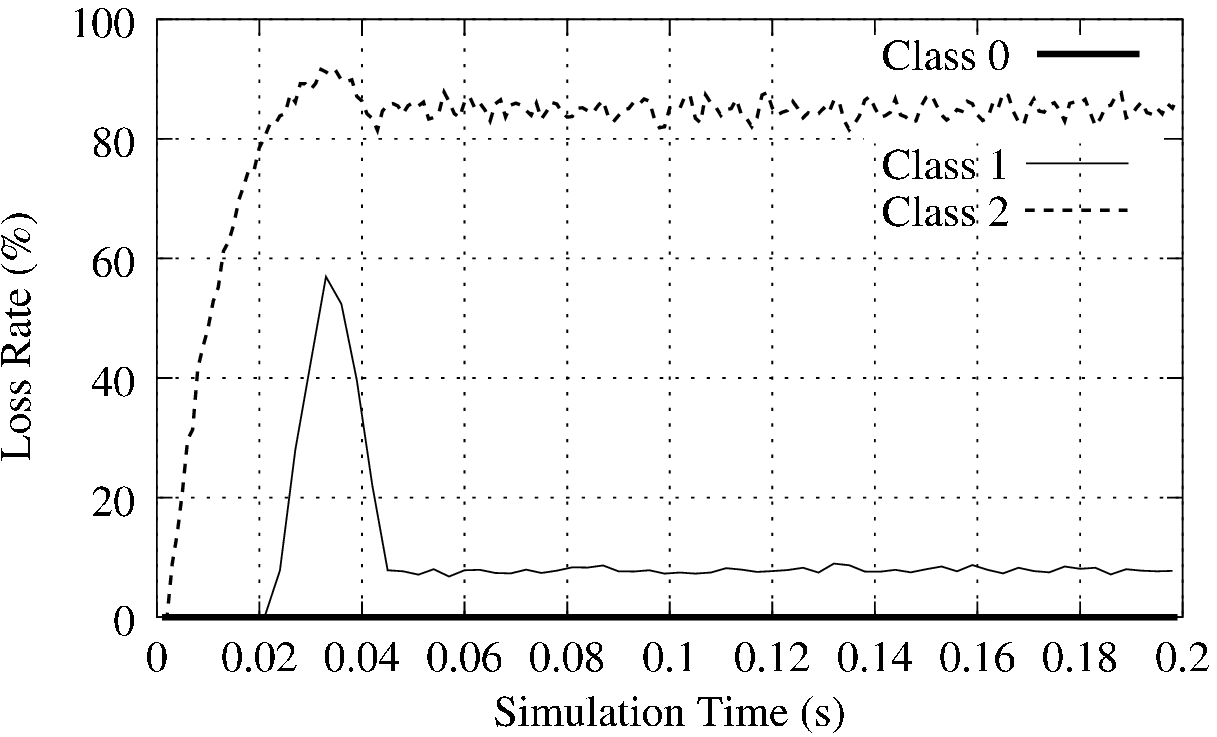}\\
        {\small (b) Input drop rate with FOQ}
}
\shortstack{
        \includegraphics[width=0.45\textwidth]{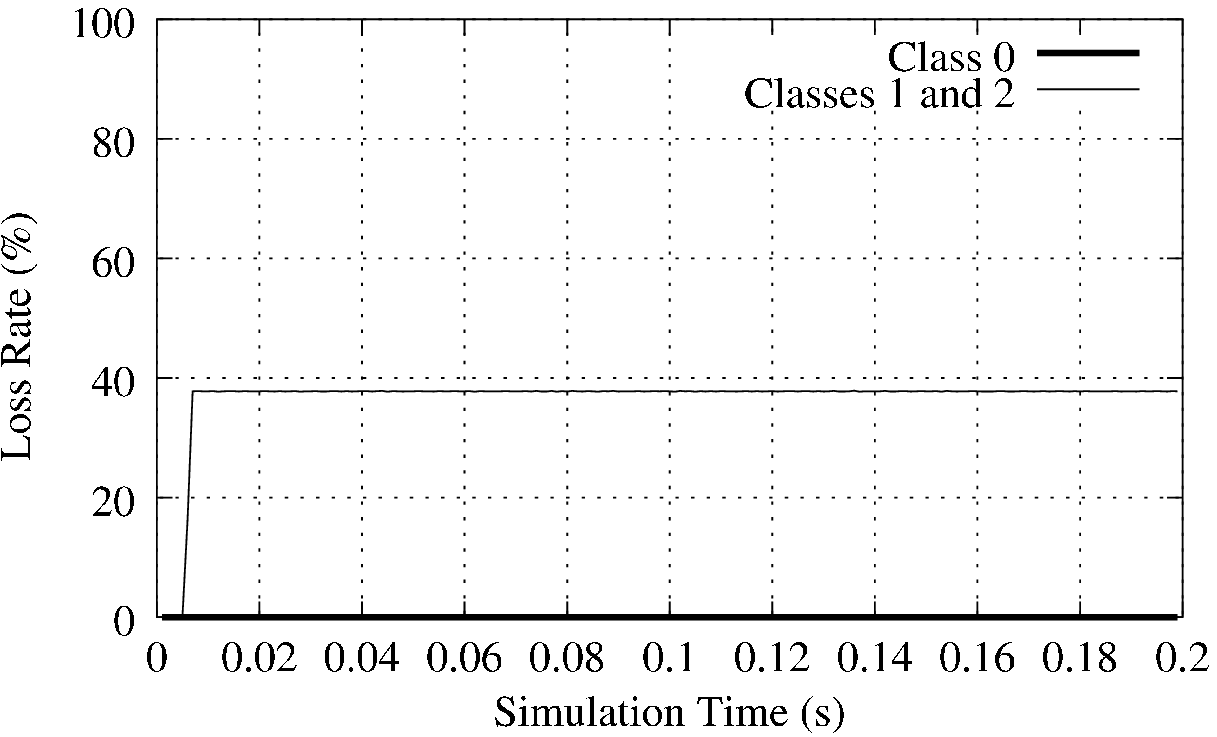}\\
        {\small (c) Fabric drop rate without FOQ}
}
\shortstack{
        \includegraphics[width=0.45\textwidth]{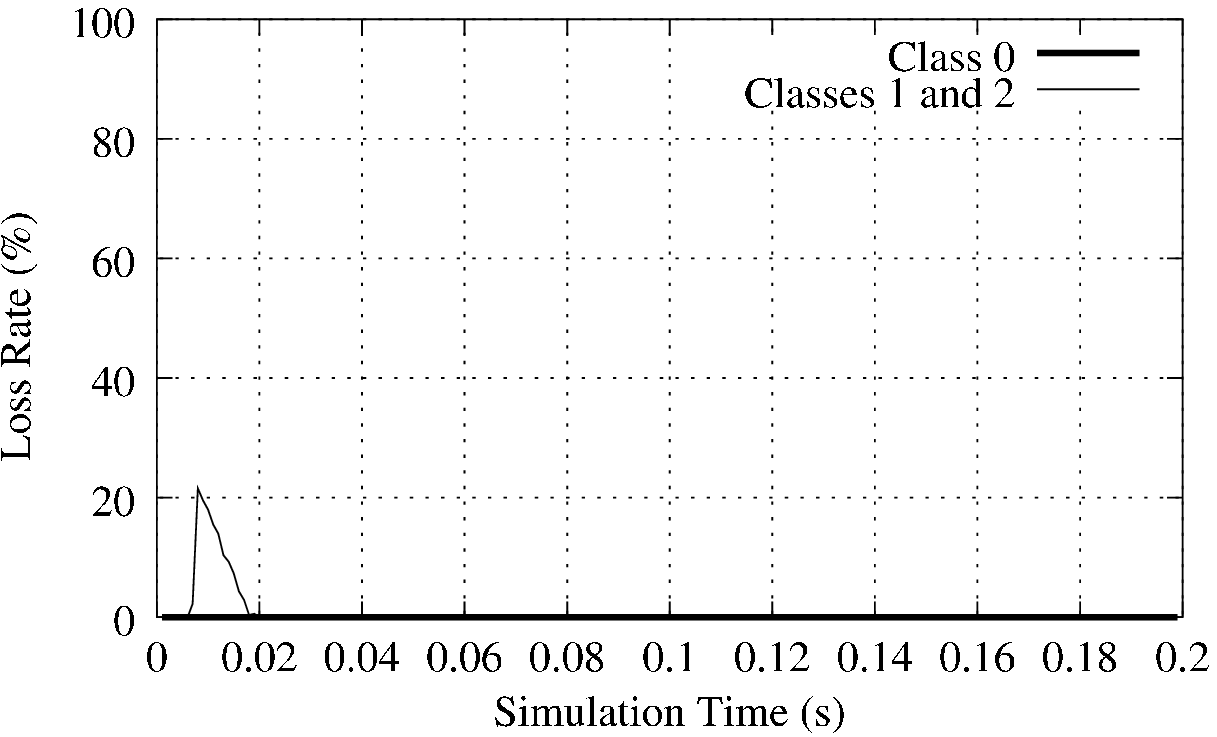}\\
        {\small (d) Fabric drop rate with FOQ}
}
\shortstack{
        \includegraphics[width=0.45\textwidth]{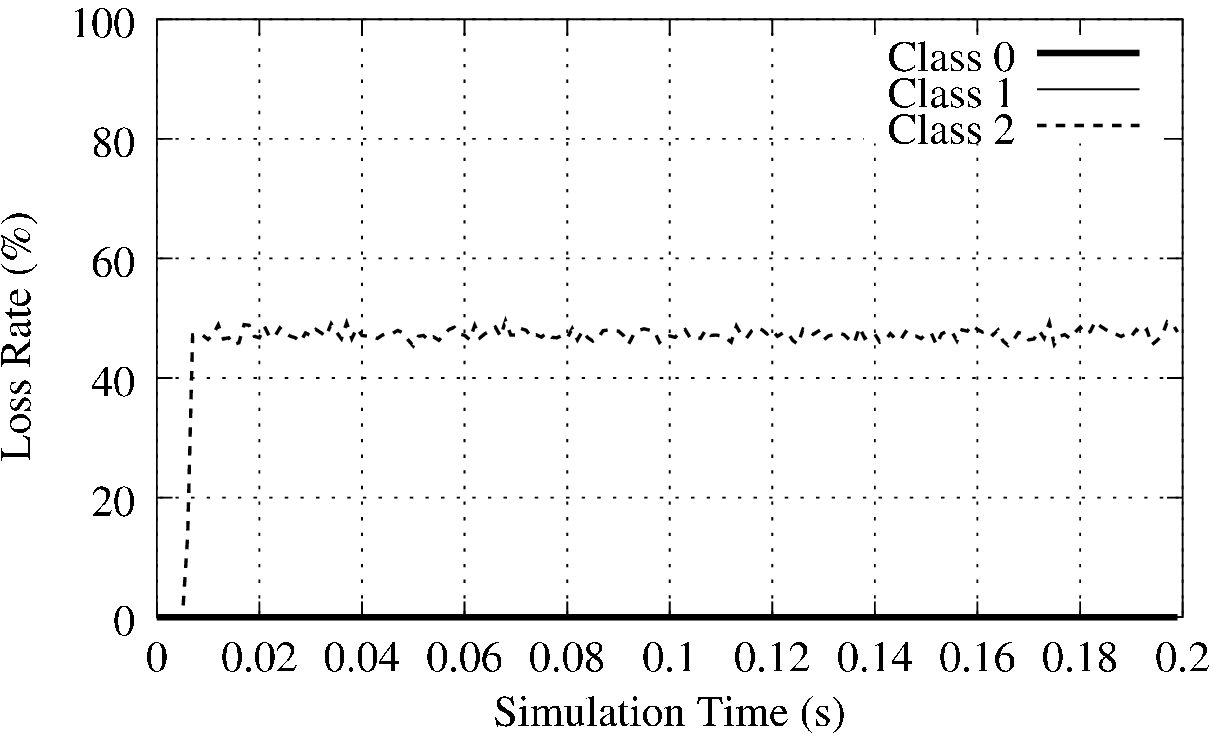}\\
        {\small (e) Output drop rate without FOQ}
}
\shortstack{
        \includegraphics[width=0.45\textwidth]{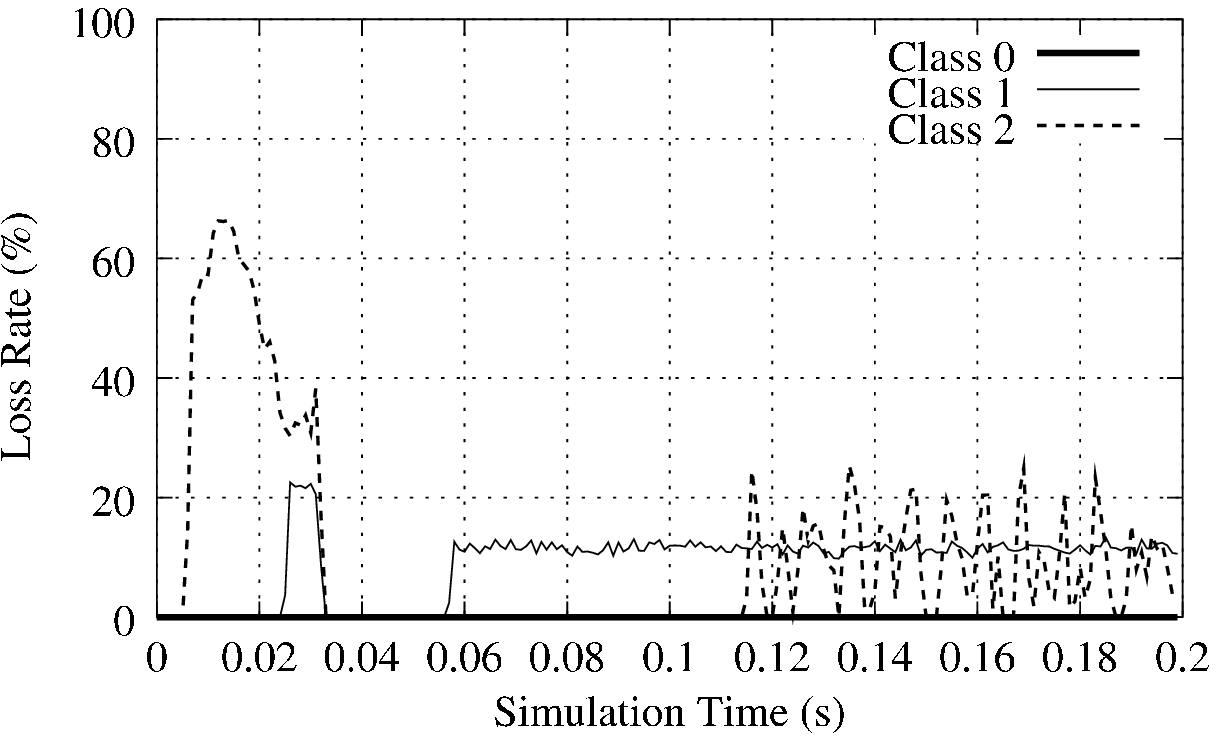}\\
        {\small (f) Output drop rate with FOQ}
}
\caption{Drop rate plots}
\label{fig-exp1-drop}
\end{center}
\end{figure*}

\begin{figure*}
\begin{center}
\shortstack{
        \includegraphics[width=0.45\textwidth]{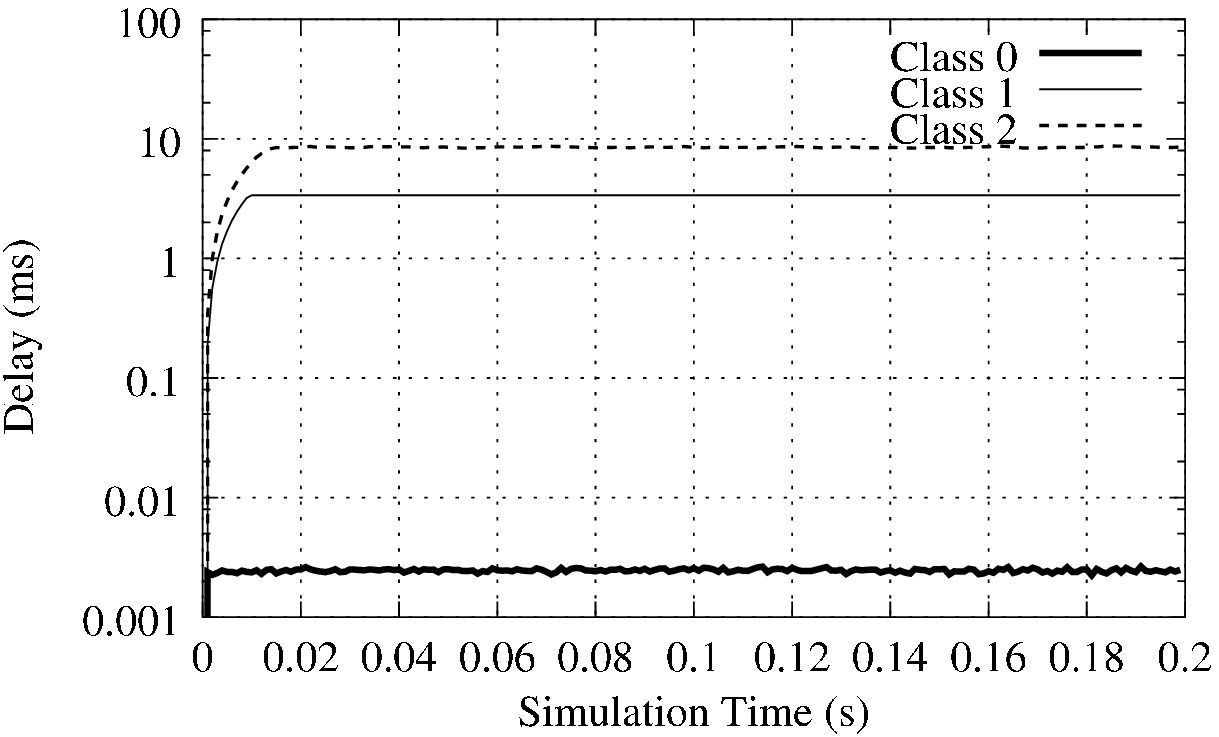}\\
        {\small (a) without FOQ}
}
\shortstack{
        \includegraphics[width=0.45\textwidth]{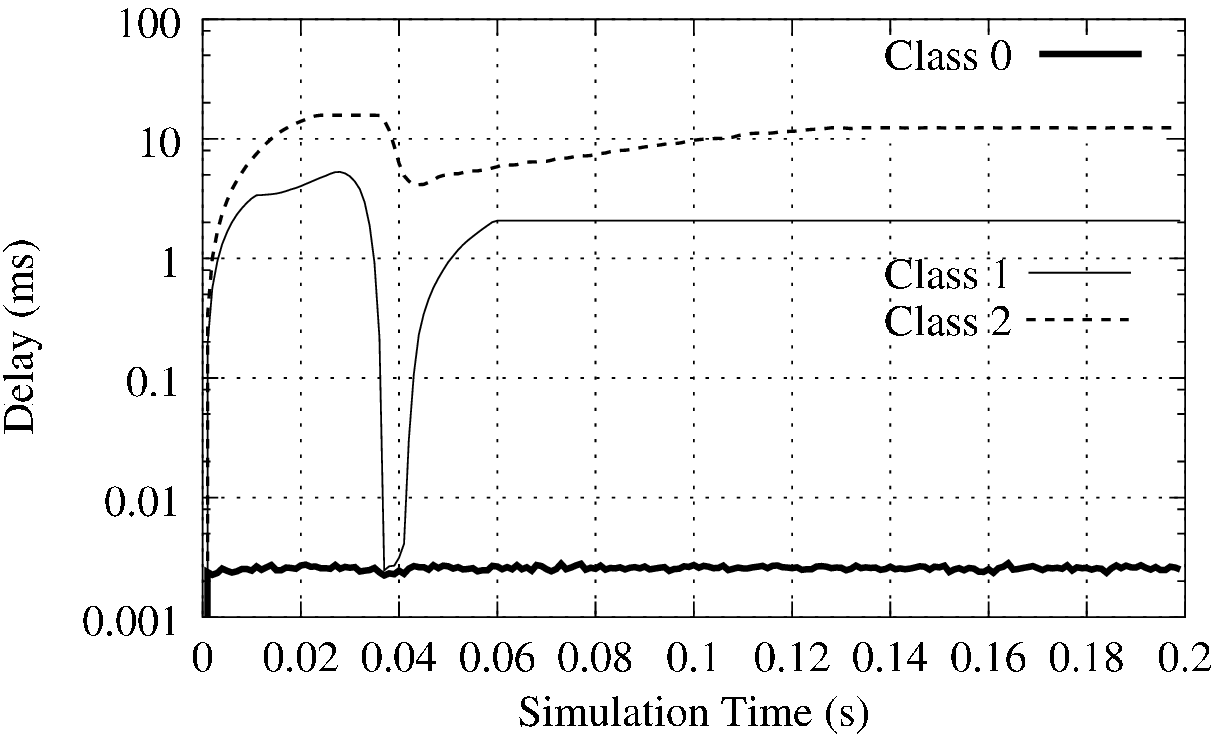}\\
        {\small (b) with FOQ}
}
\caption{Delay plots}
\label{fig-exp1-delay}
\end{center}
\end{figure*}

We simulate a 16x10~Gbps-port switch with a $5$~MB shared memory fabric
having external speedup $s=1.28$, 2~MB drop-tail OUT queues per flow,
and no ingress queues.  The FOQ-GB mechanism has a sampling rate
$T=1$~ms and feedback thresholds $d_{max}=0.17$, $d_{min}=0.02$.  We
run each simulation for $200$~ms.

The offered load is composed of three flows sending at constant rates
starting at $t=0$: flow 0: $0.952$~Gbps, flow 1 and 2: $9.52$~Gbps each,
all ingressing on separate ports and exiting the same port.  Given
that the total offered load is $20$~Gbps, the OUT port has a potential
$200$\% overload. The required guarantee for flow 0 is Premium service
($0.952$~Gbps rate guarantee), and minimum rate guarantees of
$7.75$~Gbps and $1.3$~Gbps are required for flows 1 and 2
respectively. Flow 0 is assigned to Fabric queue 0 at high priority,
and flows 2 and 3 to Fabric queue 1 at lower priority.  At the OUT
scheduler, each flow is assigned a separate queue. Queue 0 is
scheduled at high priority, whereas queues 2 and 3 are scheduled at
lower priority in a Weigted Fair Queuing discipline between them with
$6:1$ weights, corresponding to the required rate guarantees.

In Figure~\ref{fig-exp1-thru} we plot the
evolution in time of the service rate for the three flows, without and
with FOQ respectively.  In Figure~\ref{fig-exp1-drop} 
we show the dynamics of drop rate for the same
scenarios.  In all plots, each datapoint corresponds to an average over a sliding window of size 1~ms. Flow 0 is serviced at its arrival rate in both cases, due
to its high priority assignment in the fabric and OUT scheduler.  But
the rate received by flow 1 in the non-FOQ case, $5.93$~Gbps
(Figure~\ref{fig-exp1-thru}(a)), is below its requirement.  This is due
to the drop in the fabric queue 1 (Figure~\ref{fig-exp1-drop}(c)) without
discrimination between flows 1 and 2.  When using FOQ
(Figure~\ref{fig-exp1-thru}(b)), flow 1 receives $7.62$~Gbps and flow 2
$1.37$~Gbps, thus both achieving their minimum rate guarantees.  This
is explained by the FOQ action reflected in Figure~\ref{fig-exp1-drop}(b)
where we see an increase of input drop for flows 1 and 2 as a reaction
to output congestion. As a consequence, the fabric drop is zero almost
all the time in the FOQ case, in contrast with the high drop rate in
the base case. The spike in fabric drop is due to the transient state
where ingress drop is increasing but not yet sufficient for
eliminating fabric congestion.  With FOQ, fabric drop occurs only at
bursts with high rate and long duration.  It can be mitigated by
larger fabric memory or higher frequency of feedback.  Also note that
flow 0 is not affected even during the FOQ transient due to its
assignment to the high priority fabric queue.

In Figure~\ref{fig-exp1-delay} we show the
dynamics of packet transit delay through the whole switch.  While flow
0 receives minimum delay in both cases due to its high priority
assignment, flows 1 and 2 experience delays that are proportional to
their respective service rates (their OUT queues are close to full in
the steady state due to the drop-tail queue management).

\subsection{FOQ Dynamics with TCP Traffic}
\label{subsec:tcp}
Next, we examine the interaction of FOQ-GB with TCP traffic. To that
effect, we run a simulation where 4,500~TCP sources send traffic
through a switch. In this experiment, we only consider one class
of traffic. Four subnets containing 1,000~TCP sources each and one
subnet containing 500~TCP sources are connected to the switch by
five independent 1~Gbps links. All sources send traffic to the same
destination subnet, which is also connected to the switch by a 1~Gbps
link, with a one-way propagation delay of 20~ms. We have the number of
active TCP flows increase over time as follows. Each source in the first
subnet starts sending traffic between $t=0$~s and $t=1$~s, according to
a uniform random variable. Then, each source in the second subnet starts
sending traffic between $t=2$~s and $t=3$~s. Subsequently, every two
seconds, sources in an additional subnet start transmitting. Hence, we
have no overload between $t=0$~s and $t=2$~s, a potential 2:1 overload
in the fabric between $t=2$~s and $t=4$~s, a 3:1 overload between
$t=4$~s and $t=6$~s, a 4:1 overload between $t=6$~s and $t=8$~s, and
a 5:1 overload then on. There is a potential $s:1$ bottleneck at the
output port of the switch governing the 1~Gbps link to the destination
subnet after $t=2$~s. All TCP sources send 1,040-byte packets.

\begin{figure*}
\begin{center}
\shortstack{
        \includegraphics[width=0.45\textwidth]{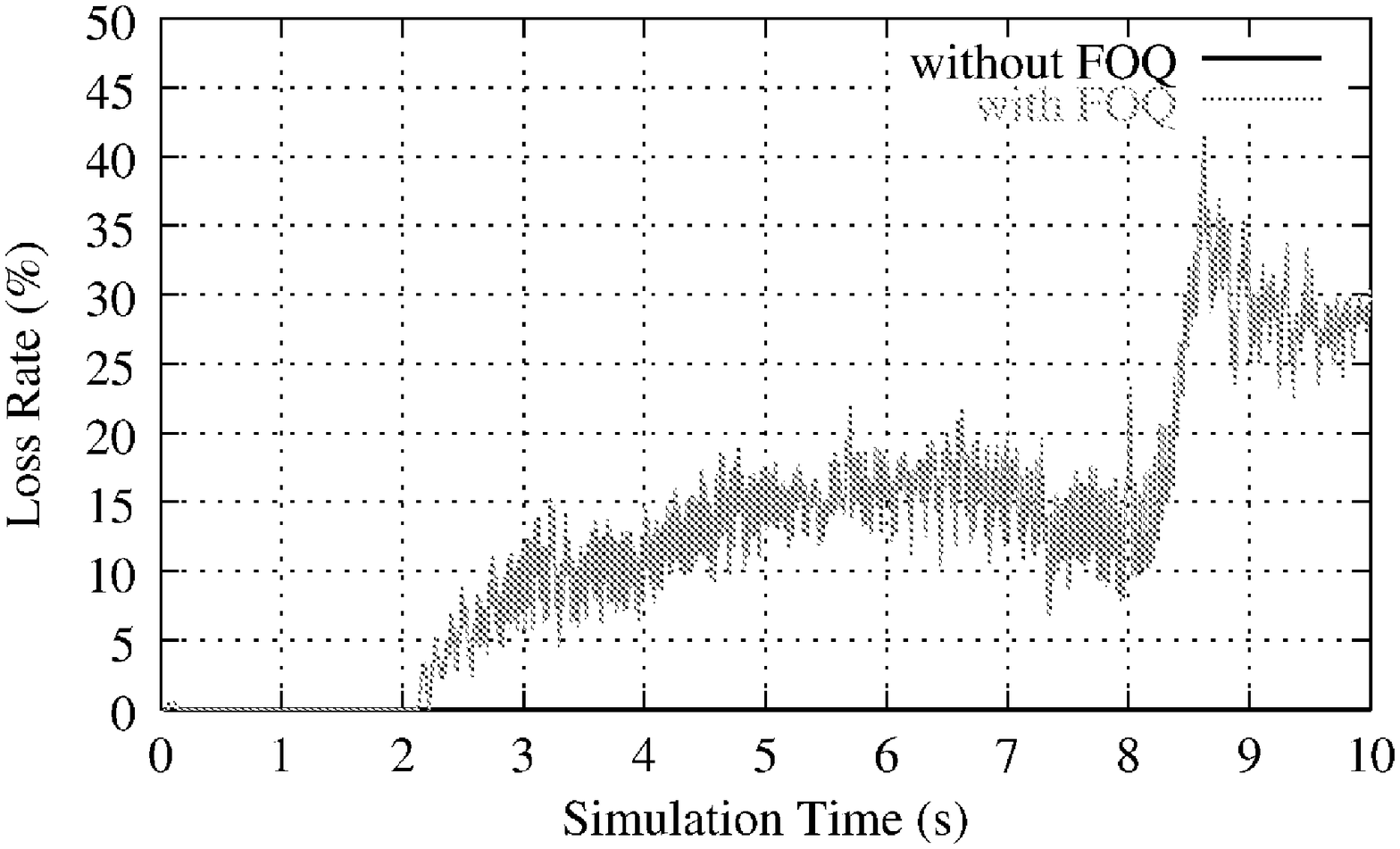}\\
        {\small (a) Input drops}
}
\shortstack{
        \includegraphics[width=0.45\textwidth]{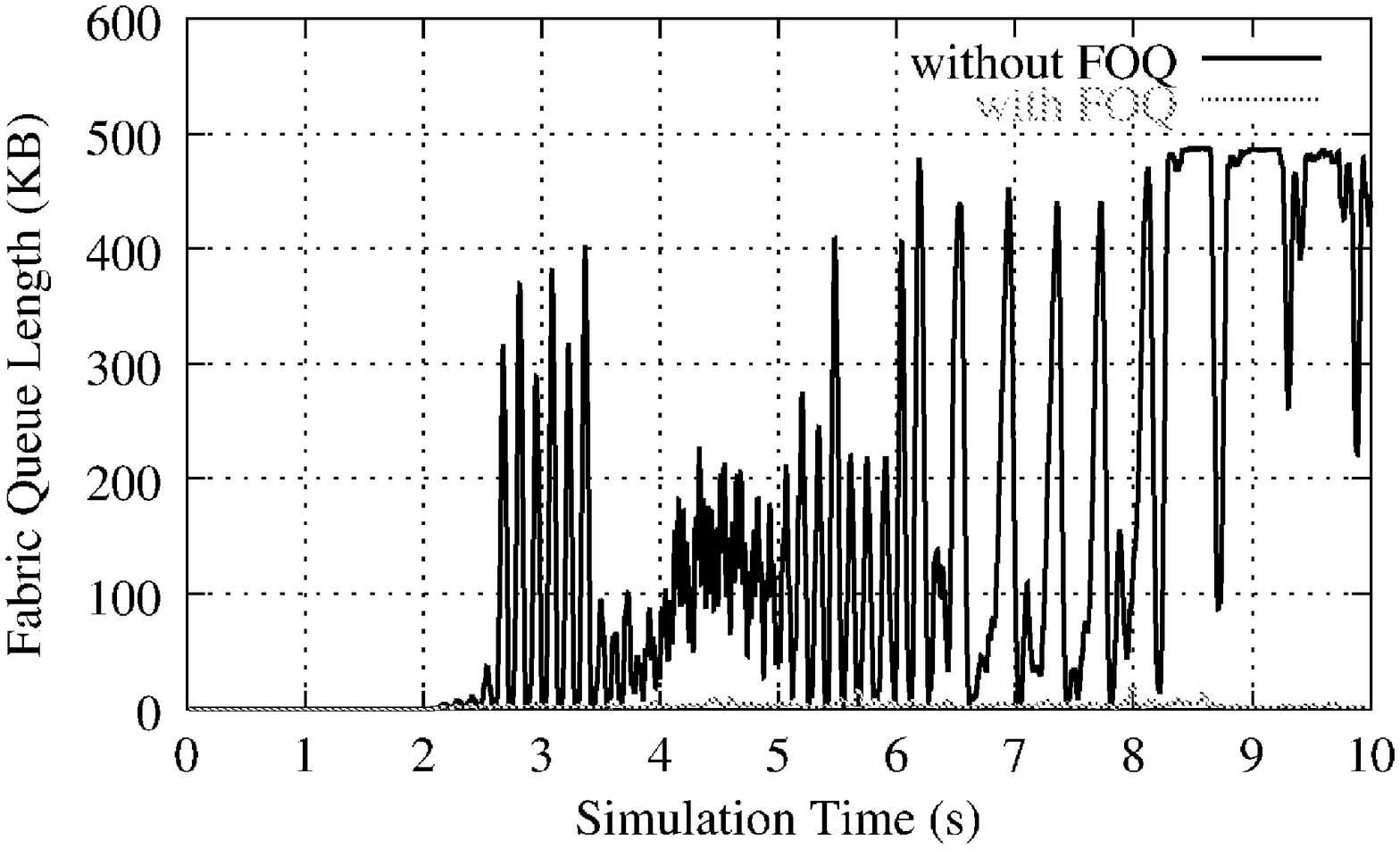}\\
        {\small (b) Fabric queue length}
}
\caption{Ingress drops and fabric queue. {\small FOQ manages to maintain a low fabric queue by dropping packets at the input links. When FOQ is not present, there are no input drops.}}
\label{fig:tcp-indrop-fabq}
\end{center}
\end{figure*}

The FOQ parameters, are chosen as in the previous experiment, i.e.,
$s=1.28$, $d_{\max} = 0.17$ and $d_{\min} = 0.02$. The fabric queue has
now a size of 500~KB and the output queue has a size of 400~KB. The
output queue runs RED, with $\max_{P} = 0.5$, $\max_{TH} = 300$~KB,
$\min_{TH} = 100$~KB, a sampling time of 1~ms, and a weight $w_q = 0.1$.
We compare the performance of the switch with and without FOQ.

\begin{figure*}
\begin{center}
\shortstack{
        \includegraphics[width=0.45\textwidth]{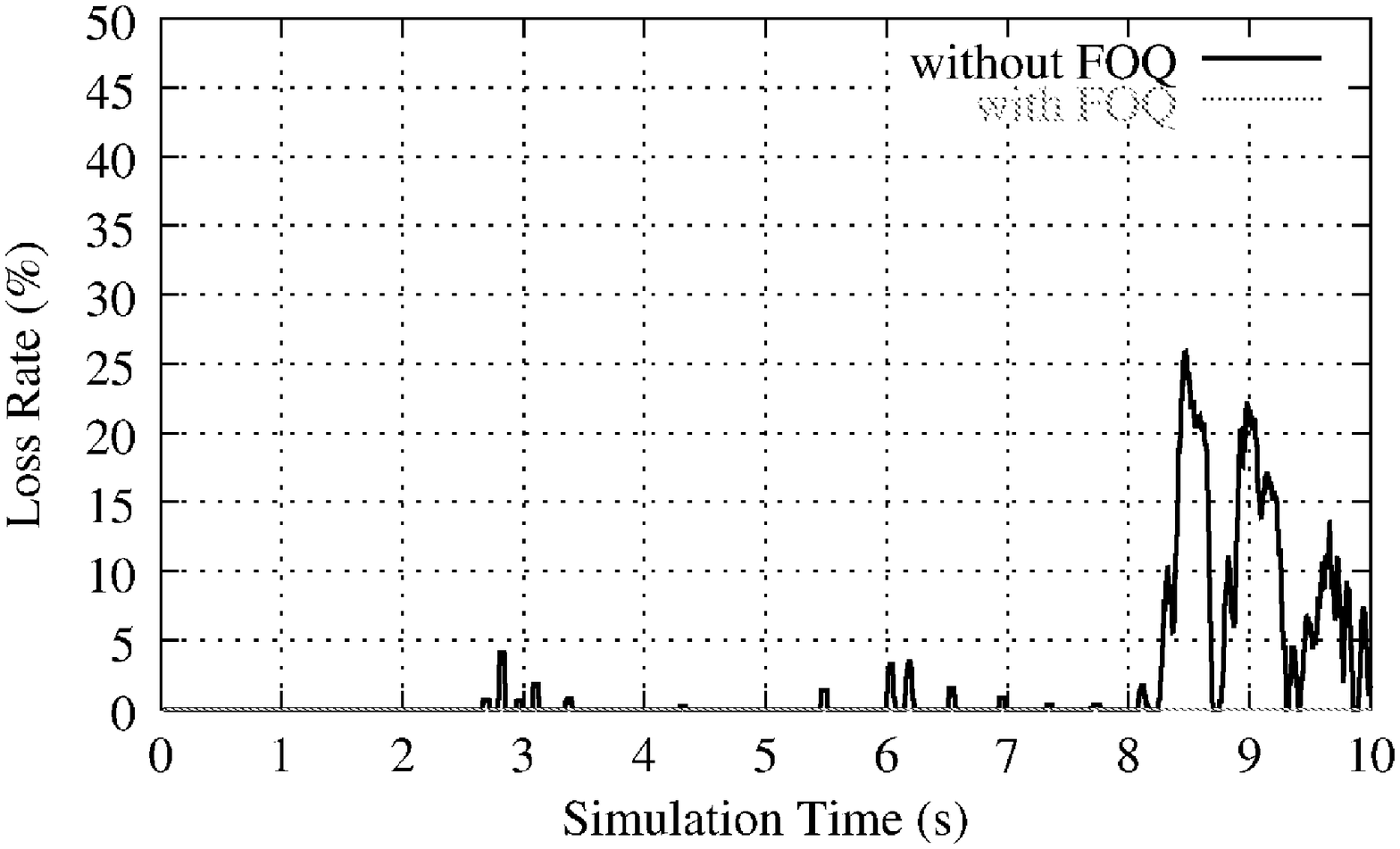}\\
        {\small (a) Fabric losses}
}
\shortstack{
        \includegraphics[width=0.45\textwidth]{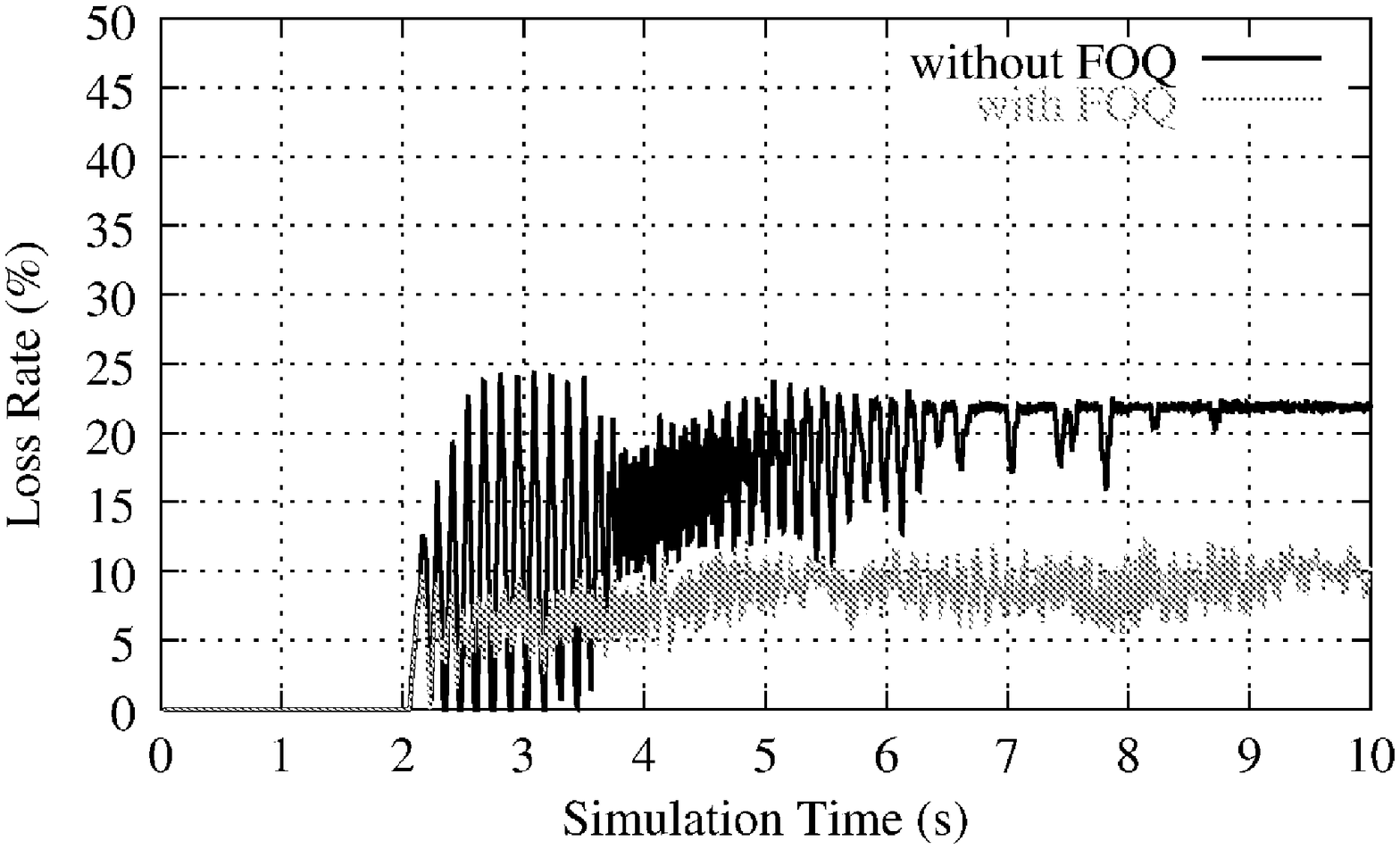}\\
        {\small (b) Output losses}
}
\caption{Fabric and output losses. {\small FOQ manages to completely avoid fabric losses, and also significantly reduces the amount of traffic dropped at the output link.}}
\label{fig:tcp-fab-egr-losses}
\end{center}
\end{figure*}

We first observe in Figure~\ref{fig:tcp-indrop-fabq}(b), where
each datapoint represents a moving average over a sliding window
of size 50~ms, that, regardless of the potential overload, FOQ
consistently manages to maintain the fabric backlog extremely close
to zero, by dropping packets at the input links. As illustrated in
Figure~\ref{fig:tcp-indrop-fabq}(a), input drops increase with the
overload. Conversely, without FOQ, and therefore in the absence of input
drops, the fabric buffer is filling up with the number of
active TCP sources, and is eventually completely full once all
sources have started transmitting. Ultimately, as illustrated in
Figure~\ref{fig:tcp-fab-egr-losses}(a), traffic is dropped in the
fabric. There are no fabric drops when FOQ is used.

\begin{figure*}
\begin{center}
\shortstack{
        \includegraphics[width=0.45\textwidth]{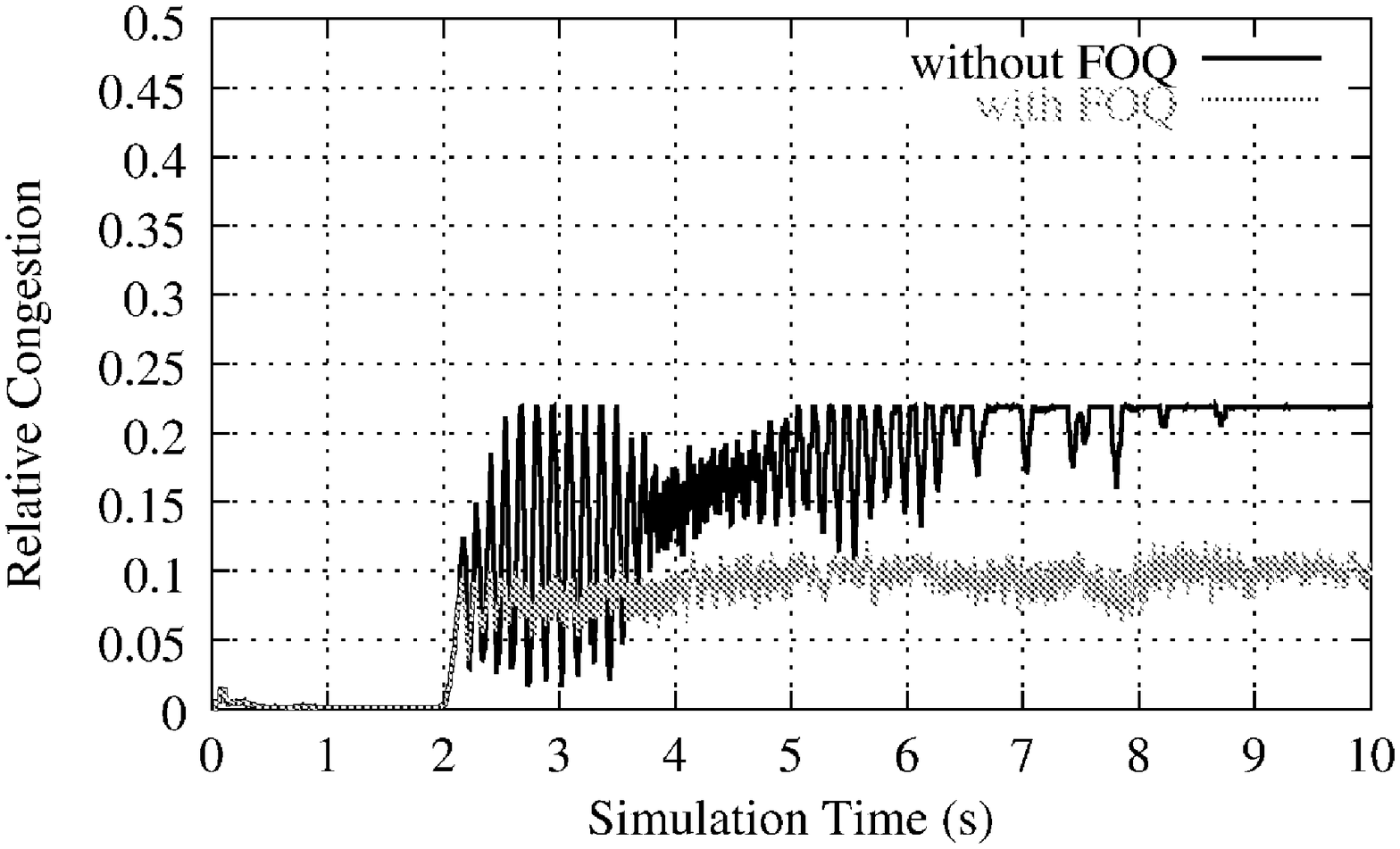}\\
        {\small (a) Egress relative congestion}
}
\shortstack{
        \includegraphics[width=0.45\textwidth]{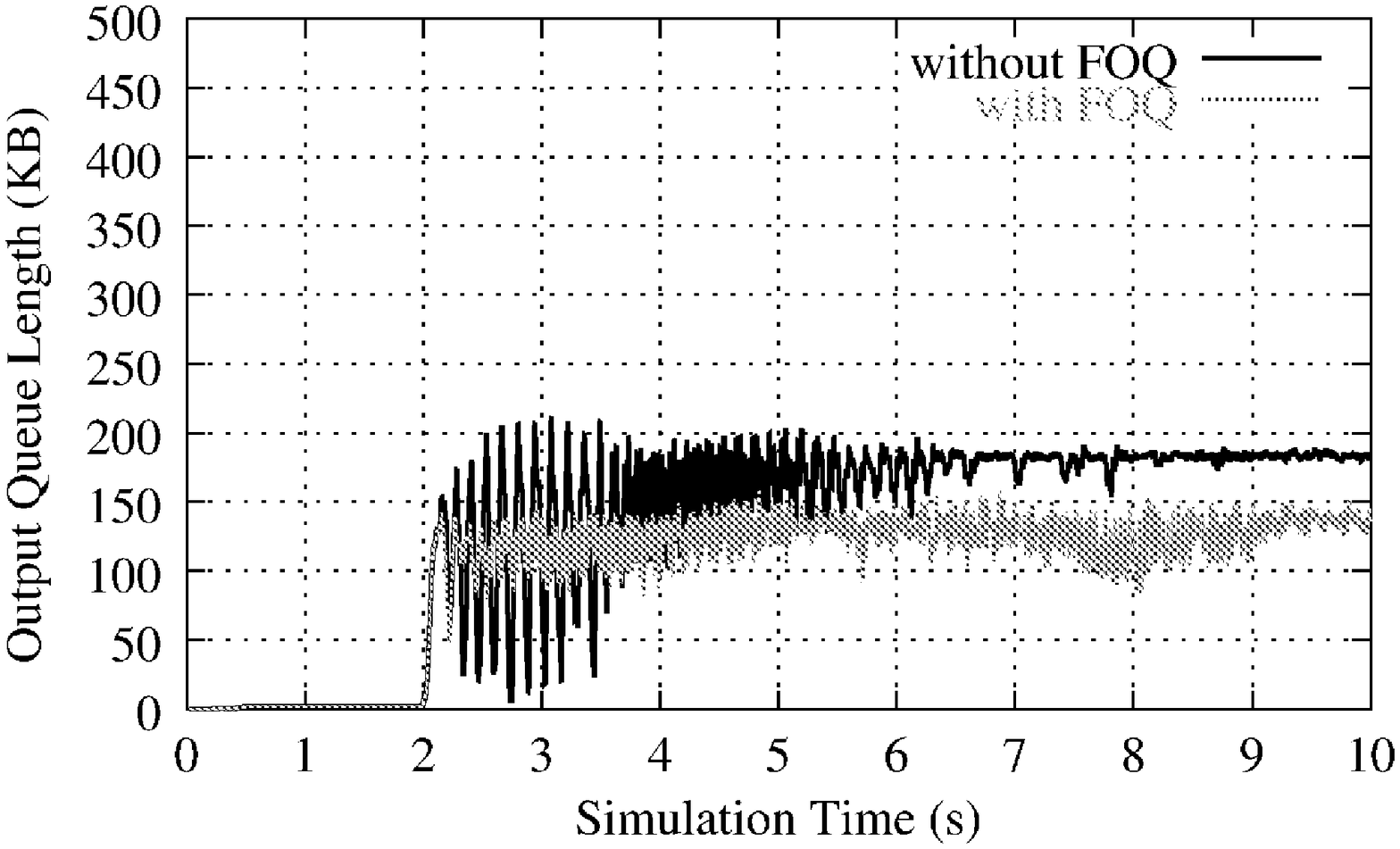}\\
        {\small (b) Output queue}
}
\caption{Relative congestion and output queue. {\small FOQ maintains the
relative congestion between $d_{\min}$ and $d_{\max}$.}}
\label{fig:tcp-relcong-outq}
\end{center}
\end{figure*}

Last, we observe in Figure~\ref{fig:tcp-fab-egr-losses}(b) that
the output loss rate is limited by $1-1/s \approx 21.8$\% when FOQ is disabled.
On the other hand, FOQ maintains the egress relative congestion
close to $d_{\mbox{\scriptsize mid}} = 0.098$, as shown in
Figure~\ref{fig:tcp-relcong-outq}(a), and consequently, the
output loss rate remains close to 9.8\%. When the loss rates
become roughly constant, the output queue length, represented in
Figure~\ref{fig:tcp-relcong-outq}(b), also becomes constant, by
virtue of a stable RED control \cite{Firoiu99-RED-Infocom}.

As a conclusion to this second experiment, we have shown that FOQ's
objectives of preventing fabric drops and regulating the traffic that
arrives at the output link were met in the case of an experiment with a
large number of TCP sources. The results were even more positive than
those obtained with constant-rate sources, as FOQ does not exhibit
transient behaviors in this scenario. This can be justified by the fact
that FOQ feedback is run at a much higher frequency (every $T=1$~ms)
than the TCP congestion control algorithms, which are run with an
approximately 40-ms delay here.

\section{Discussion and Conclusions} 
\label{sec-conclusion}
In this paper we presented the Feedback Output Queuing
architecture for packet switching that provides support for service 
guarantees 
when the switching speed is limited by the memory
read and write speeds. Using a fast switching fabric in this case
leads to a build-up in fabric buffers and eventually either to buffer
overflow and packet discarding or to unbounded delays at the fabric
inputs due to backpressure. The FOQ architecture solves this problem
by triggering packet discard only from flows that exceed their
allocated bandwidth, and therefore limiting the build-up and delay at
the fabric buffers. In the worst case the arrival rate will be
$\lambda_{max}$, the total input capacity of the fabric. For the PI
controller the maximum fabric queue size and the maximum delay in the
fabric can be calculated from (\ref{queuesize}) by inserting
$\lambda_0=\lambda_{max}$. Any delay value above this number can be
deterministically guaranteed to a flow by using a proper scheduler
(e.g. WFQ-based) at the output queues after the fabric.

An alternative approach to solve the same problem is to use VOQ at
fabric inputs. Recent studies show that VOQ can also provide
deterministic delay bounds
\cite{Nong-providingQoS-Globecomm-1999}. This is however at the
expense of computational complexity.  VOQ algorithms require $O(N^2)$
computations per packet slot to determine which packets will be sent
to their destinations. This high computational complexity makes the
VOQ approach less feasible for high bit-rate switches. In contrast,
the FOQ requires a total of $O(N)$ computations per packet slot and
$O(KN)$ computations per feedback interval, where $K$ is the number of
supported classes.  Since the feedback interval is much larger than a
packet slot, computations for the feedback are actually
negligible. Furthermore, the computations are distributed to the
inputs and outputs, so that each input and output performs $O(1)$
computations. In other words, FOQ's computational
complexity is much lower than VOQ, the current state of the art.

We applied discrete feedback control theory to derive a stable
configuration for FOQ.  Through analysis and simulations we showed that a
quantized version of a PI controller named ``Gear-Box control'' is
stable, responds quickly to traffic bursts and provides highly
accurate QoS guarantees.  

We believe that this work has sparked many venues for future research.
There is a range of control algorithms to be investigated besides
those presented here.  The interaction between the TCP congestion
control algorithm and FOQ (and RED queue management) is an interesing
control problem.  The FOQ architecture can be extended with a set of
input queues in order to provide zero loss for a wider range of bursty
traffic, given a limited fabric memory size.

\section*{Acknowledgments}

The authors would like to thank Eric Haversat, Tom Holtey and Franco
Travostino of Nortel Networks for many useful discussions.

\bibliographystyle{plain}

\begin{thebibliography}{10}

\bibitem{Anderson-hispeedswitch-tocs-1993}
T.~Anderson, S.~Owicki, J.~Saxe, and C.~Thacker.
\newblock High speed switch scheduling for local area networks.
\newblock {\em ACM Transactions on Computer Systems}, 11(4):319--352, November
  1993.

\bibitem{Waldemar-VPLS-req-IETF}
W.~Augustyn, G.~Heron, V.~Kompella, M.~Lassere, P.~Menezes, H.~Ould-Brahim, and
  T.~Senevirathne.
\newblock {Requirements for Virtual Private LAN Services (VPLS)}.
\newblock IETF draft, draft-ietf-l2vpn-vpls-requirements-00.txt, October 2002.

\bibitem{dsarch-rfc}
S.~Blake, D.~Black, M.~Carlson, E.~Davies, Z.~Wang, and W.~Weiss.
\newblock An architecture for differentiated services.
\newblock IETF RFC 2475, December 1998.

\bibitem{Carugi-PPVPNreq-2002}
M.~Carugi, D.~McDysan, L.~Fang, F.~Johansson, A.~Nagarajan, J.~Sumimoto, and
  R.~Wilder.
\newblock Service requirements for layer 3 provider provisioned virtual private
  networks.
\newblock IETF draft, draft-ietf-ppvpn-requirements-04.txt, March 2002.

\bibitem{McKeown99a}
S.-T. Chuang, A.~Goel, N.~McKeown, and B.~Prabhakar.
\newblock Matching output queueing with a combined input-output queued switch.
\newblock In {\em Proceedings of IEEE INFOCOM '99}, volume~3, pages 1169--1178,
  New York, NY, March 1999.

\bibitem{Davie02-EFPHB-IETF}
B.~Davie, A.~Charny, J.~Bennett, K.~Benson, J.-Y. {Le Boudec}, W.~Courtney,
  S.~Davari, V.~Firoiu, and D.~Stiliadis.
\newblock An expedited forwarding {PHB}.
\newblock IETF RFC 3246, March 2002.

\bibitem{Firoiu99-RED-Infocom}
V.~Firoiu and M.~Borden.
\newblock A study of active queue management for congestion control.
\newblock In {\em Proceedings of {IEEE INFOCOM'00}}, volume~3, pages
  1435--1444, Tel-Aviv, Israel, April 2000.

\bibitem{Firoiu02}
V.~Firoiu, X.~Zhang, and E.~G{\"u}nd{\"u}zhan.
\newblock Feedback output queueing: a novel architecture for efficient
  switching systems.
\newblock In {\em Proceedings of Hot Interconnects X}, Stanford, CA, August
  2002.

\bibitem{Floyd-red}
S.~Floyd and V.~Jacobson.
\newblock {R}andom early detection for congestion avoidance.
\newblock {\em IEEE/ACM Transactions on Networking}, 1(4):397--413, July 1993.

\bibitem{Franklin-digitalcontrol-1998}
G.~Franklin, J.~Powell, and M.~Workman.
\newblock {\em Digital control of dynamic systems}.
\newblock {Addison-Wesley}, Menlo Park, CA, 3rd edition, 1998.

\bibitem{Gleeson00-VPN-framework-IETF}
B.~Gleeson, A.~Lin, J.~Heinanen, G.~Armitage, and A.~Malis.
\newblock A framework for {IP} based virtual private networks.
\newblock IETF RFC 2764, February 2000.

\bibitem{Goderis02-SLS-IETF}
D.~Goderis, S.~Van~Den Bosch, Y.~T'joens, O.~Poupel, C.~Jacquenet, G.~Memenios,
  G.~Pavlou, R.~Egan, D.~Griffin, P.~Georgatsos~L. Georgiadis, and P.~Van
  Heuven.
\newblock Service level specification semantics and parameters.
\newblock IETF draft, draft-tequila-sls-02.txt, February 2002.

\bibitem{GuerinPla01}
R.~Gu\'erin and V.~Pla.
\newblock Aggregation and conformance in differentiated service networks: A
  case study.
\newblock {\em ACM Computer Communication Review}, 31(1):21--32, January 2001.

\bibitem{Heinanen-AFPHB-1999}
J.~Heinanen, F.~Baker, W.~Weiss, and J.~Wroclawski.
\newblock Assured forwarding {PHB} group.
\newblock IETF RFC 2597, June 1999.

\bibitem{Jacobson:cong-avoid}
V.~Jacobson.
\newblock Congestion avoidance and control.
\newblock In {\em {Proceedings of ACM SIGCOMM'88}}, pages 314--329, Stanford,
  CA, August 1988.

\bibitem{Karr00-reduced-HotI}
K.~Kar, T.V. Lakshman, D.~Stiliadis, and L.~Tassiulas.
\newblock Reduced complexity input buffered switches.
\newblock In {\em Proceedings of Hot Interconnects VIII}, Stanford, CA, August
  2000.

\bibitem{Creary00-trends-CAIDA}
S.~McCreary and K.~Claffy.
\newblock {Trends in Wide Area IP Traffic Patterns}.
\newblock CAIDA, May 2000.

\bibitem{McKeown-comparison-CNIS-1998}
N.~McKeown and T.~Anderson.
\newblock A quantitative comparison of iterative scheduling algorithms for
  input-queued switches.
\newblock {\em Computer Networks and ISDN Systems}, 30(24):2309--2326, December
  1998.

\bibitem{Prizma}
C.~Minkenberg and T.~Engbersen.
\newblock A combined input and output queued packet-switched system based on
  {Prizma} switch-on-a-chip technology.
\newblock {\em IEEE Communications Magazine}, 38(12):70--77, December 2000.

\bibitem{Nong-provision-commmag-2000}
G.~Nong and M.~Hamdi.
\newblock On the provisioning of {Quality of Service} guarantees for input
  queued switches.
\newblock {\em IEEE Communications Magazine}, 38(12):62--69, December 2000.

\bibitem{Nong-providingQoS-Globecomm-1999}
G.~Nong and M.~Hamdi.
\newblock Providing {QoS} guarantees for unicast/multicast traffic with fixed
  and variable-length packets in multiple input-queued switches.
\newblock In {\em Proceedings of IEEE ISCC'01}, pages 166--171, 2001.

\bibitem{TCPchar-TON-00}
J.~Padhye, V.~Firoiu, D.~Towsley, and J.~Kurose.
\newblock {Modeling TCP Reno Performance: A Simple Model and Its Empirical
  Validation}.
\newblock {\em IEEE/ACM Transactions on Networking}, 8(2):133--145, April 2000.

\bibitem{Rosen-L2VPN-2001}
E.~Rosen, C.~Filsfils, G.~Heron, A.~Malis, L.~Martini, and S.~Vogelsang.
\newblock An architecture for {L2VPNs}.
\newblock IETF draft, draft-ietf-ppvpn-l2vpn-00.txt, July 2001.

\bibitem{Stevens-rfc2001}
W.~Stevens.
\newblock {TCP} slow start, congestion avoidance, fast retransmit, and fast
  recovery algorithms.
\newblock IETF RFC 2001, January 1997.

\bibitem{Xu-individual-2002}
Y.~Xu and R.~Guerin.
\newblock Individual {QoS} versus aggregate {QoS}: A loss performance study.
\newblock In {\em Proceedings of IEEE INFOCOM '02}, volume~3, pages 1170 --
  1179, New York, NY, June 2002.
\end{thebibliography}

\newpage
\appendix
\section*{Appendix}
In this appendix we give a detailed derivation of some of the equations.

Taking the $z$-transforms of (\ref{out_rate}) and (\ref{rhonew}),  we get
\begin{eqnarray}
\nonumber
\rho(z) & = & K\left(R(z)-R_{opt}(z)\right)\\
\nonumber
 & &+K_I\frac{z}{z-1}\left( R(z)-R_{opt}(z)\right)\\
\label{rhoz1}
 & &+S_{N_0}(z)
\end{eqnarray}
and
\begin{equation}
R(z)=\lambda(z)-z^{-1}\rho (z).
\label{Rz1}
\end{equation}
Transfer functions of this system between the output rate, $R$, and
the two inputs and initial state, $\lambda$, $R_{opt}$, and $S_{N_0}$, are given by
$$
\frac{R(z)}{\lambda(z)}=\frac{z(z-1)}{z^2+(K+K_I-1)z-K},
$$
$$
\frac{R(z)}{R_{opt}(z)}=\frac{(K+K_I)z-K}{z^2+(K+K_I-1)z-K},
$$
and
$$
\frac{R(z)}{S_{N_0}(z)}=\frac{1-z}{z^2+(K+K_I-1)z-K}.
$$
Let $z_1$ and $z_2$ be two roots of the system characteristic equation, i.e.
$$
z_{1,2}^2+(K+K_I-1)z_{1,2}-K=0.
$$
Then without loss of generality
$$
z_1=-\frac{K+K_I-1}{2}+\frac{1}{2}\sqrt{(K+K_I-1)^2+4K}
$$
$$
z_2=-\frac{K+K_I-1}{2}-\frac{1}{2}\sqrt{(K+K_I-1)^2+4K}.
$$
We showed in Proposition 1 that the system is stable if
$$
0<K_I<2(1-K).
$$
We next find the solution for the drop rate $\rho$ assuming this stability condition
is satisfied. For step inputs and initial condition, $\lambda(z)=z\lambda /(z-1)$,
$R_{opt}(z)=zr_{opt}/(z-1)$, $S_{N_0}(z)=zS_{N_0}/(z-1)$, and defining
$D=\lambda-r_{opt}$ as the difference between the arrival and the desired
rates, we have from (\ref{rhoz1}) and (\ref{Rz1}):
\begin{eqnarray}
\rho(z)  = &\frac{KD\frac{z}{z-1}+
K_ID\frac{z^2}{(z-1)^2}+S_{N_0}
\frac{z}{z-1}}
{1+\frac{K}{z}+\frac{K_I}{z-1}}\nonumber\\
 =& z^2\frac{[(K+K_I)D+S_{N_0}]z-KD-S_{N_0}}
{(z-1)(z^2+(K+K_I-1)z-K)}.\nonumber
\end{eqnarray}
This can be written as a partial fraction expansion as
$$
\rho(z)=D\left(\frac{z}{z-1}-\frac{A_1z}{z-z_1}+\frac{A_2z}{z-z_2}\right)
$$
where
$$
A_1=\frac{z_1^2-\frac{S_{N_0}}{D}z_1}{z_1-z_2}
$$
and
$$
A_2=\frac{z_2^2-\frac{S_{N_0}}{D}z_2}{z_1-z_2},
$$
which can be solved easily. Finally recall that this system was obtained initially
by defining a new time axis for $n\geq N_0$. Therefore after taking the inverse
$z$-transform we combine the result with $n<N_0$ case to get
$$
\rho[n]=\left\{\begin{array}{ll}
[K+(n+1)K_I](sc-r_{opt}), & n< N_0 \\
D(1-A_1z_1^{n-N_0}+A_2z_2^{n-N_0}), & n\geq N_0 \end{array}\right. .
$$
\end{document}